\documentclass[10pt]{article}
\usepackage{amsmath}
\usepackage{graphicx}
\usepackage{color}
\usepackage{hyperref}
\hypersetup{colorlinks=true, linkcolor=blue, citecolor=blue, urlcolor=blue}
\usepackage[table,xcdraw]{xcolor}
\usepackage[caption=false]{subfig}
\usepackage{capt-of}
\begin{document}
\baselineskip=16pt

\begin{center}
{\LARGE  Ay\'{o}n-Beato-Garc\'{i}a black hole in AdS space-time surrounded by quintessence: geodesic, shadow and thermodynamics  }
\end{center}

\vspace{0.2cm}

\begin{center}
{\bf Ahmad Al-Badawi}\footnote{\bf email: ahmadbadawi@ahu.edu.jo}\\ 
{\it Department of Physics, Al-Hussein Bin Talal University, 71111, Ma'an, Jordan}\\
\vspace{0.1cm}
{\bf Faizuddin Ahmed}\footnote{\bf email: faizuddinahmed15@gmail.com (Corresponding author)}\\ 
{\it Department of Physics, University of Science \& Technology Meghalaya, Ri-Bhoi, Meghalaya, 793101, India}\\
\vspace{0.1cm}
{\bf \.{I}zzet Sakall{\i}}\footnote{\bf email: izzet.sakalli@emu.edu.tr}\\ 
{\it Department of Physics, Eastern Mediterranean University, 99628, Famagusta Northern Cyprus, via Mersin 10, Turkiye}
\end{center}

\vspace{0.2cm}

\begin{abstract}
In this study, we present a novel exact solution to the gravitational field equations, known as the Ayón-Beato-García black hole solution, set against the backdrop of anti-de Sitter space and surrounded by a quintessence field. This solution serves as an interpolation between three distinct anti-de Sitter black hole configurations, namely, the Ayón-Beato-García, Schwarzschild-Kiselev, and the standard Schwarzschild black hole solutions. The first aspect of our investigation focuses on the geodesic motion of particles, where we explore how the black hole's space-time geometry-incorporating the effects of nonlinear electrodynamics, the quintessence field, and the curvature radius-influences the dynamics of both massless and massive particles near the black hole. To further enrich our analysis, we extend the study to include the perturbative dynamics of a massless scalar field within the black hole solution, placing special emphasis on the scalar perturbative potential. Subsequently, we focus into the phenomenon of black hole shadows, examining how various parameters, such as the nonlinear electrodynamics, the curvature of space-time and the presence of the quintessence field, impact the size and shape of the shadow cast by the black hole. In the final segment of our study, we shift our attention to the thermodynamics of the black hole solution. We compute several essential thermodynamic quantities, including the Hawking temperature, specific heat capacity, and Gibbs free energy, analyzing how these properties evolve in response to changes in the various parameters that define the space-time geometry, which in turn affect the gravitational field when compared to the traditional black hole solutions.
\end{abstract}


\section{Introduction}\label{sec:1}

Several regular black hole (BH) solutions have been discovered through the coupling of gravity with non-linear electrodynamics (NLED) theories. The pioneering work in this field was initiated by Born and Infeld, who introduced NLED to address the singularities associated with point charges and mitigate the energy divergence \cite{MB1}. Later, Peres provided a comprehensive discussion of NLED in the context of general relativity \cite{AP1}. This framework gained further importance when NLED was shown to emerge as a limiting case in specific string theory models \cite{NS1}. It is well-established that at high energies, the linearity of classical electrodynamics breaks down due to interactions with other physical fields, making NLED a natural and promising alternative for describing electromagnetic phenomena in such extreme conditions. In these high-energy regimes, NLED theories offer new insights by serving as a source of gravity, capable of generating a diverse range of new BH solutions.

One notable contribution in this direction was the development of a regular black hole solution sourced by NLED, which satisfied the weak energy condition (WEC) and asymptotically approached the Reissner-Nordström black hole \cite{EAB1}. This work laid the foundation for further advancements, as subsequent studies derived additional regular black hole solutions by coupling NLED with Einstein's gravity \cite{EAB2, EAB5}. Later, it was demonstrated that Bardeen's regular black hole could be interpreted as a solution arising from NED \cite{EAB3}. Motivated by this work, several researchers endeavored to construct black hole solutions in general relativity with NLED as the source (see, Refs. \cite{EAB4, KE1, BP1, DVS1, PP1, YM1}). Additionally, various families of regular black hole metrics sourced by NLED have been explored in the literature \cite{LB1}, expanding the range of viable models for regular black holes in NLED theories \cite{LB2}. These solutions significantly deviate from traditional black hole models, with profound implications not only for black hole physics but also for cosmology and string theory. Indeed, NLED has found applications in both cosmological contexts \cite{VADL1, MN1, RPM1} and string theory \cite{NS1, ESF1}, reinforcing its versatility and relevance in modern theoretical physics. 

This investigation aims to study the effects of both NLED and quintessence fields (QFs) on black hole geometry analogue to the case for cosmic string (CS), discussed in \cite{AK1, AK2}. The interplay between these two factors may reveal new aspects of gravitational dynamics, offering insights into the formation, evolution, and structure of black holes, as well as their broader cosmological implications. The primary motivation for this study is to contribute to our understanding of the gravitational effects induced by these fundamental cosmic objects and their role in shaping the universe. The first aspect of our study explores the geodesic motion of test particles, probing how the BH's spacetime geometry, including the ABG parameter, the QF, and the AdS curvature, influences the trajectories of both massless and massive particles \cite{isz01,isz02,isz03}. We analyze the structure of the effective potential for particle motion and investigate the conditions for stable and unstable orbits, showing the intricate relationship between the BH's parameters and the dynamics of test particles \cite{isz04,isz05,isz06,isz07}. Furthermore, we extend our analysis to the perturbative dynamics of a spin-0 massless scalar field in this ABG BH solution, focusing on the scalar perturbative potential \cite{isz08,isz09,isz10}. Here, we examine in detail how the geometry of the spacetime, including the ABG parameter, the QF constant, and the radius of curvature, modifies the characteristics of the potential and affects the propagation of scalar fields in the BH background \cite{isz11,isz12,isz13}. We then study the BH shadow in this setup, investigating how various parameters-such as the ABG parameter, QF constant, and the AdS curvature-affect the size and shape of the shadow \cite{isz14,isz15,isz16,isz17}. This analysis provides important insights into the observable features of the BH and demonstrates how deviations from standard BH models can alter the appearance of the BH shadow in astrophysical observations \cite{isz18,isz19,isz20,isz21}. In the final part of the study, we shift our focus to the thermodynamics of the ABG BH solution \cite{isz22,isz23,isz24,isz242,isz243}. We compute key thermodynamic quantities, including the Hawking temperature, specific heat capacity, and Gibbs free energy, and examine how these thermodynamic properties are influenced by the ABG parameter, the QF, and the AdS space curvature \cite{isz25,isz26,isz262,isz263,isz27,isz28}. This comprehensive thermodynamic analysis is compared to the standard BH solutions found in the literature, providing new insights into the modification of BH thermodynamics in the presence of ABG parameter, QFs, and AdS backgrounds \cite{isz29,isz30,isz31,isz32}. The combined results of this study offer a deeper understanding of the intricate interplay between gravitational dynamics, scalar field perturbations, BH observables, and thermodynamics in the context of modified gravitational models, such as those involving QFs and AdS backgrounds \cite{isz33,isz34,isz35,isz36}. In summary, our motivation for studying ABG BHs in AdS spacetime with quintessence stems from their unique role as theoretical laboratories for probing the interface between general relativity and quantum theory \cite{isz37,isz38}. These regular BH models elegantly resolve the central singularity problem while preserving key classical BH features \cite{isz39}. The analysis of geodesic structure, shadows, and thermodynamics of these solutions reveal observable signatures testable with current or future astrophysical instruments, potentially advancing our understanding of both quantum gravity and cosmological evolution \cite{isz40}.

This paper is structured as follows: In Section \ref{sec:2}, we present a static and spherically symmetric Ayan-Betao-Garcia BH solution in an AdS space surrounded by a QF. Section \ref{sec:3} provides a detailed analysis of the geodesic motion of both massless and massive particles. Section \ref{sec:4} focuses on the dynamics of a spin-0 massless scalar field, governed by the Klein-Gordon equation, and derives the Schrödinger-like wave equation. In Section \ref{sec:5}, we examine the BH shadow and its characteristics. In Section \ref{sec:6}, we investigate the thermal properties of the BH solution, exploring how various parameters inherent in the BH's spacetime geometry-such as the ABG parameter, the QF, and the AdS curvature-affect the phenomena under study. Specifically, we analyze the impact of these parameters on the motion of test particles, the propagation of scalar fields, the size and shape of the BH shadow, and the thermodynamic properties of the BH. In Section \ref{sec:7}, we summarize our findings and present our conclusions.

\section{New ABG BH solution with Quintessence}\label{sec:2}

The action describing Einstein's gravity coupled to ABG NLED and surrounded by a QF in 4 dimensions is given by
\begin{equation}
S=\int d^{4}x\sqrt{-\Tilde{g} }\left[ \frac{1}{2\kappa }\left( R-2\Lambda \right) -%
\frac{1}{4\pi }\mathcal{L}_{ABG}(F)+\mathcal{L}^{q}\right] 
\end{equation}%
where $\Lambda $ is the cosmological constant  linked to
 the AdS length ($\Lambda=-3/ \ell^2_p$),  $\Tilde{g}$ is determinant of metric and  
$R$ is Ricci curvature scalar.  $\mathcal{L}_{ABG}(F)$ and $\mathcal{L}^{q}$ describe
Lagrangian density for ABG NLED sources and the Lagrangian of the quintessence dark energy,  respectively. The Lagrangian $\mathcal{L}_{ABG}(F)$ is function of $F=\frac{1}{4}%
F_{\mu \nu }F^{\mu \nu }$, where $F_{\mu \nu }=\nabla _{\mu }A_{\nu }-\nabla
_{\nu }A_{\mu }$ is the electromagnetic field strength tensor which is
associated with the gauge potential $A_{\mu }.$

The Lagrangian density $\mathcal{L}_{ABG}(F)$  \cite{ayon,beato,ayon2,ayon3,ayon4}
\begin{equation}
\mathcal{L}_{ABG}(F)=\frac{F\left(1-3 \sqrt{2g^{2}F }\right)}{\left(1+ \sqrt{2g^{2}F}\right)
^{3}}-\frac{3M}{g^3}                             
     \left( \frac{\left( 2g^{2}F\right) ^{5/4}}{\left(1+ \sqrt{2g^{2}F}\right)
^{5/2}}\right),
\label{ne11}
\end{equation}%
where, $g$ and $M$ correspond to the ABG parameter and the mass of the BH, respectively. \\ The Lagrangian for the QF is \cite{MER1} 
\begin{equation}
\mathcal{L}^{q}=-\frac{6wc}{8\pi }\left( \frac{2F}{g^{2}}\right) ^{\frac{3}{4}\left(
w+1\right) },
\end{equation}%
where $c$ is the normalization constant related to the density of
quintessence $\rho _{q}=-\frac{3cw}{2r^{3w+3}}$ and $w$ is the state parameter with 
$-1 < w < -1/3$.  The stress-energy tensor has
the following components \cite{VVK}
\[
T_{t}^{t}=T_{r}^{r}=\rho _{q},
\]%
\begin{equation}
T_{\theta }^{\theta }=T_{\phi }^{\phi }=-\frac{1}{2}\rho _{q}\left(
1+3w\right) .  \label{qu12}
\end{equation}

Variation of the action with respect to the metric $\Tilde{g}_{\mu \nu }$ yields equations of motion. 
\begin{equation}
G_{\mu \nu }+\Lambda g_{\mu \nu }=T_{\mu \nu }^{ABG}+T_{\mu
\nu }^{q},  \label{eom11}
\end{equation}%
\begin{equation}
\nabla _{\mu }\left( \frac{\partial L^{NE}}{\partial F}F^{\mu \nu }\right)
=0,\quad \nabla _{\mu }\left( \ast F^{\mu \nu }\right) =0,
\end{equation}%
where the Einstein tensor $G_{\mu \nu }=R_{\mu \nu }-\frac{1}{2}g_{\mu \nu }R$  and $T_{\mu \nu }^{ABG}$ and $
T_{\mu \nu }^{q}$ are the energy-momentum tensor related to the ABG NLED and
QF, respectively. The energy-momentum tensor associated to
ABG NLED is given by%
\begin{equation}
T_{\mu \nu }^{ABG}=2\frac{\partial \mathcal{L}_{ABG}}{\partial F}F_{\mu \rho }F_{\nu
}^{\rho }-2\Tilde{g}_{\mu \nu }\mathcal{L}_{ABG}  \label{emt12}
\end{equation}
To determine the BH solution with ABG NLED sources surrounding with QF, we consider the following static spherically symmetric spacetime line element in 4-dimensions spacetime is given by
\begin{equation}\label{metric}
ds^{2}=-f\left( r\right) dt^{2}+\frac{dr^{2}}{f\left( r\right) }+r^{2}\left(
d\theta ^{2}+\sin ^{2}\theta d\phi ^{2}\right) ,
\end{equation}%
where 
\begin{equation}
f\left( r\right) =1-\frac{2m\left( r\right) }{r}.
\end{equation}%
First, we examine
Maxwell's field strength tensor $F_{\mu \nu }$ using the following magnetic
charge choice
\begin{equation}
F_{\mu \nu }=2\delta _{\lbrack \mu }^{\theta }\delta _{\nu ]}^{\phi }Z\left(
r,\theta \right) .
\end{equation}%
Following \cite{ayon2}, we obtain 
\begin{equation}
F_{\mu \nu }=2\delta _{\lbrack \mu }^{\theta }\delta _{\nu ]}^{\phi }g\left(
r\right) \sin \theta .
\end{equation}
which leads to $g(r)=g$, where we choose the integration constant as $g$ and it is identified as the  MM charge:\begin{equation}
    \frac{1}{4\pi}\int_S
Fds=\frac{g}{4\pi}\int_0 ^\pi\int_0 ^{2\pi} \sin\theta d\theta d \phi =g,\end{equation} where $S$ is the spherical surface at infinity. Therefore, the non-vanishing component of $F_{\mu \nu }$ is $F_{\theta \phi
}=g\left( r\right) \sin \theta $ and potential $A_{\phi }=-g\left( r\right)
\cos \theta $ \cite{ayon2}. Using $dF=0,$ we obtain $g(r)=g=const.$, where $g$ is the
MM charge. Hence, the magnetic field strength is given by 
\begin{equation}
F_{\theta \phi }=g\sin \theta ,\quad F=\frac{g^{2}}{2r^{4}}.  \label{com12}
\end{equation}%
Inserting Eq. (\ref{com12}) into Eq. (\ref{ne11}), then the Lagrangian
density of ABG NLED sources becomes 
\begin{equation}
    \mathcal{L}_{ABG}(F)=\frac{g^2(r^2-3g^2)}{2(r^2+g^2)^3}+\frac{8Mg^2}{(r^2+g^2)^{5/2}}.
\end{equation}
Next, we obtain the components of energy momentum tensor as 
\begin{equation}
T_{t}^{ABG\,t}=T_{r}^{ABG\,r}=\frac{g^2(r^2-3g^2)}{(r^2+g^2)^3}-\frac{6Mg^2}{(r^2+g^2)^{5/2}} .
\end{equation}
Thus, the $\left( r,r\right) $  components of Eq. (\ref{eom11}) becomes%
\begin{equation}
m^{\prime }\left( r\right) =-\frac{3r^{2}}{2\ell^{2}_p}+\frac{g^2r^2(r^2-3g^2)}{2(r^2+g^2)^3}-\frac{3Mg^2r^2}{(r^2+g^2)^{5/2}} -\frac{%
3wc }{2 r^{3w+1}}.
\end{equation}%
Integrate the above equation we obtain%
\begin{equation}
m\left( r\right) =-\frac{r^{3}}{2l^{2}_p}+\frac{M\,r^3}{\left(r^2+g^2\right)^{3 / 2}}-\frac{g^2\,r^3}{2\left(r^2+g^2\right)^2}+
\frac{c}{ 2 r^{3w}}.
\end{equation}%
Finally,  the metric function for a 4 dimensions BH with AdS ABG NLED sources in the
presence of QF is
\begin{equation}
f(r) =1-\frac{2M\,r^2}{\left(r^2+g^2\right)^{3 / 2}}+\frac{g^2\,r^2}{\left(r^2+g^2\right)^2}+\frac{r^2}{\ell^2_{p}}-\frac{c}{r^{3\,w+1}}. \label{m41}
\end{equation}

The derived regular BH solution (\ref{m41}) is characterized by the mass $M$, cosmological constant $\ell_{p} =\sqrt{-3/\Lambda}$, QF parameters $(c,w)$ and the  MM charge $g$.   The function (\ref{m41}), satisfies several limits as boundary conditions. For instance, without QF ($c=0$) results in the AdS ABG BH \cite{abg21}, setting $g=0$ and $\ell_{p} \to \infty$ results in the  Schwarzschild BH metric with quintessence \cite{VVK}. Similarly, by choosing $c=0$ and $g=0$, we obtain Schwarzschild-AdS BH metric \cite{BB6}.

In the current study, we focus on the state parameter $w=-2/3$, and hence, the metric function $f(r)$ from Eq. (\ref{m41}) reduces as 
\begin{equation}
f(r) =1-\frac{2\,M\,r^2}{\left(r^2+g^2\right)^{3 / 2}}+\frac{g^2\,r^2}{\left(r^2+g^2\right)^2}+\frac{r^2}{\ell^2_{p}}-c\,r \label{m1}
\end{equation}
with the line element describing a static and spherically symmetric ABG BH solution in AdS space background is given by the metric (\ref{metric}) as
\begin{equation}
ds^2=-f(r)\,dt^2+\frac{dr^2}{f(r)}+r^2\,(d\theta^2+\sin^2 \theta\, d\phi^2).\label{metric2}
\end{equation}

Below, we examine the geodesic motion of both massless and massive particles, the dynamics of a massless scalar field governed by the Klein-Gordon equation, the BH shadow, and the thermal properties of the BH solution (\ref{metric2}). We investigate how various parameters inherent in the BH spacetime geometry, such as the ABG parameter, the QF, and the AdS curvature, influence these phenomena. 

\section{Dynamics of massless and massive particles}\label{sec:3}

In this section, dynamics of test particles in the selected ABG BH with QF, described by the metric (\ref{metric}) with the metric function (\ref{m1}). We analyze the behavior of particles and show parameters, such as, ABG NLED parameter $g$, the radius of curvature $\ell_p$, and the QF parameter $c$ for the chosen specific state parameter $w$ influences the dynamics of test particles. Since the spacetime (\ref{metric}) is static and spherically symmetric, we can, without loss of generality, restrict our analysis of geodesics motion in the equatorial plane, where $\theta = \pi/2$ and $\dot{\theta}=0$. The Lagrangian density function, derived using the metric tensor $g_{\mu\nu}$, is given by $\mathcal{L} = \frac{1}{2}\,g_{\mu\nu}\,\dot{x}^{\mu}\,\dot{x}^{\nu}$ \cite{NPB,CJPHY,IJGMMP,AHEP4,AHEP5,EPJC,PDU,BC}, where the dot represents the derivative with respect to the affine parameter $\tau$. Using the metric (\ref{metric}), this Lagrangian density function can be explicitly expressed as follows:
\begin{equation}
 \mathcal{L}=\frac{1}{2}\,\left(-f(r)\,\dot{t}^2+\frac{\dot{r}^2}{f(r)}+r^2\,\dot{\phi}^2\right),\label{pp1}
\end{equation}

There are two Killing vectors $\xi_{(t)}$ and $\xi_{(\phi)}$ corresponds to the coordinates $(t, \phi)$ since the metric tensor $g_{\mu\nu}$ for the line-element (\ref{metric}) is independent of these. Obviously there are two constants of motions, namely energy $\mathrm{E}$ and the conserved angular momentum $\mathrm{L}$ and these are given by
\begin{eqnarray}
    &&\mathrm{E}=f(r)\,\dot{t},\label{pp2}\\
    &&\mathrm{L}=r^2\,\dot{\phi}.\label{pp3}
\end{eqnarray}

With these, we find geodesics equation for the radial coordinate $r$ as follows:
\begin{eqnarray}
    \dot{r}^2+V_\text{eff}(r)=\mathrm{E}^2,\label{pp4}
\end{eqnarray}
where the effective potential $V_\text{eff}(r)$ for null ($\kappa=0$) or time-like ($\kappa=-1$) geodesics is given by
\begin{eqnarray}
    V_\text{eff}(r)=\left(-\kappa+\frac{\mathrm{L}^2}{r^2}\right)\,\left(1-\frac{2\,M\,r^2}{\left(r^2+g^2\right)^{3 / 2}}+\frac{g^2\,r^2}{\left(r^2+g^2\right)^2}+\frac{r^2}{\ell^2_{p}}-c\,r\right).\label{pp5}
\end{eqnarray}

From the above equation (\ref{pp5}), it is evident that several factors such as ABG NLED parameter $g$, the radius of curvature $\ell_p$, and the QF parameter $c$ for the chosen specific state parameter $w=2/3$ influences the effective potential of the system and modified the results. Moreover, the presence of QF alters the effective potential compared to the known results in the literature.

\begin{figure}[ht!]
    \centering
    \subfloat[$c=0.01$]{\centering{}\includegraphics[width=0.4\linewidth]{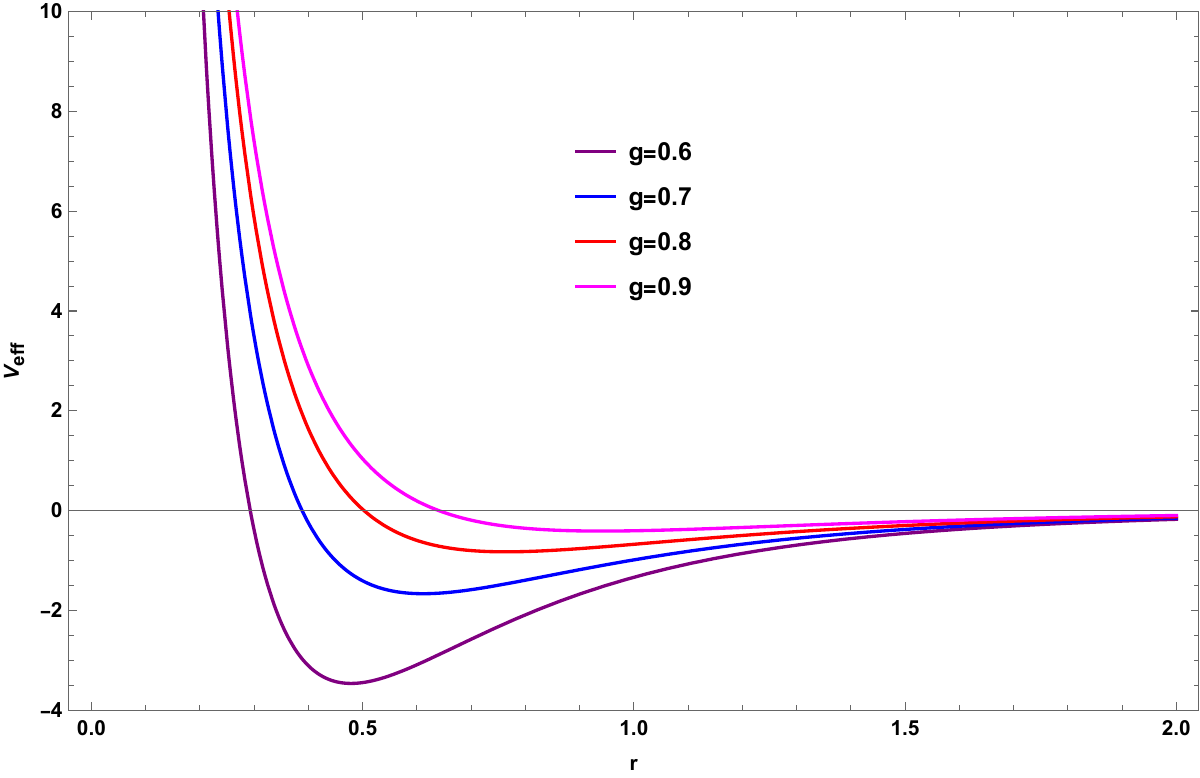}}\quad\quad\quad
    \subfloat[$g=0.9$]{\centering{}\includegraphics[width=0.4\linewidth]{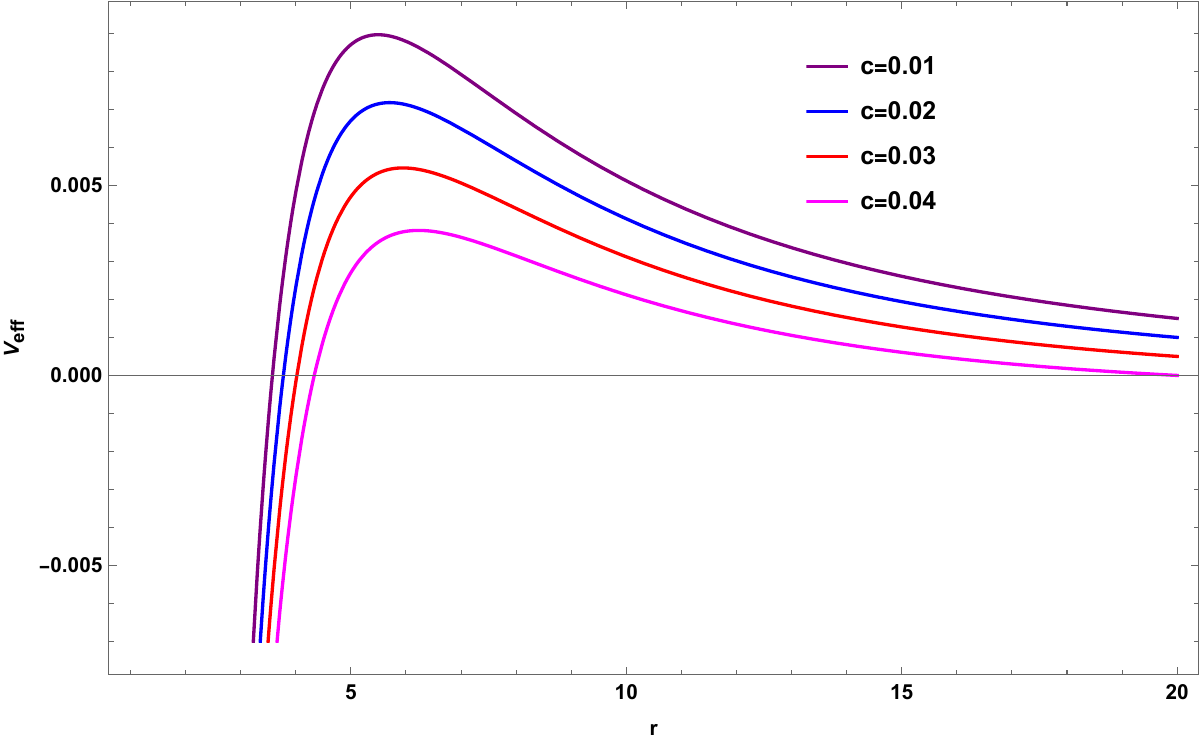}}\\
    \subfloat[]{\centering{}\includegraphics[width=0.4\linewidth]{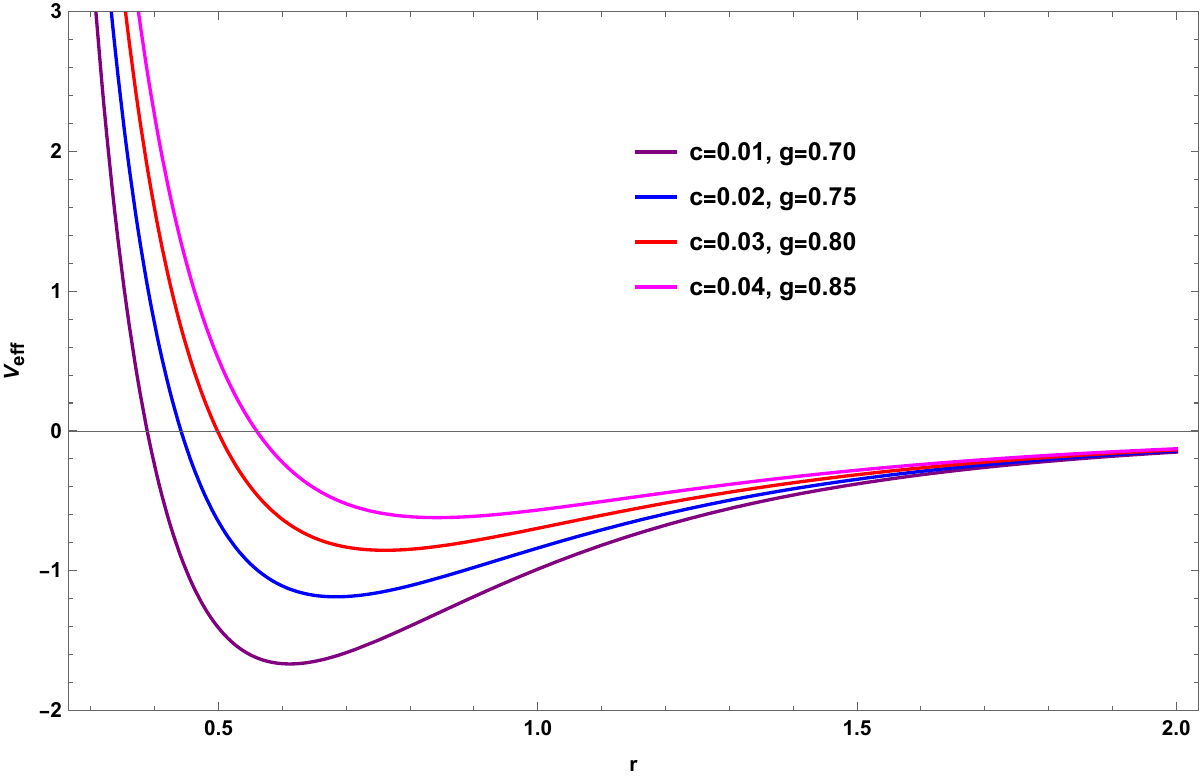}}
    \caption{The behavior of the effective potential $V_\text{eff}$ for null geodesics for different values of $g$ and $c$. Here, we set $M=2$ and $\mathrm{L}=1$}
    \label{fig:null-potential}
\end{figure}

\begin{figure}[ht!]
    \centering
    \includegraphics[width=0.45\linewidth]{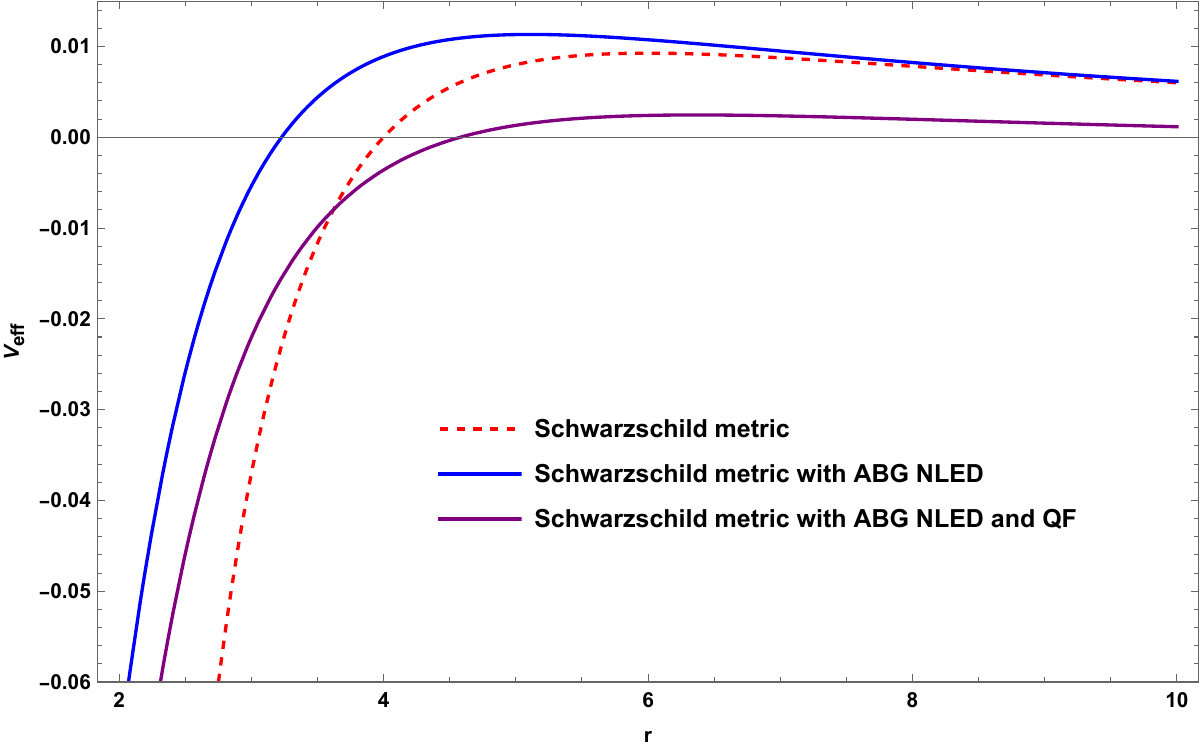}
    \caption{A comparison of the effective potential $V_\text{eff}$ for null geodesics for different BHs. Here, we set $g=1$, $c=0.05$, $M=2$ and $\mathrm{L}=1$.}
    \label{fig:null-comparison}
\end{figure}

\subsection{Optical Properties}

In this part, we study optical properties for the selected BH metric and discuss how the ABG NLED parameter $g$, the radius of curvature $\ell_p$, and the QF parameter $c$ for the chosen state parameter $w=2/3$ influences the properties. For light-like geodesics, $\kappa=0$, the effective potential from Eq. (\ref{pp5}) becomes
\begin{eqnarray}
    V_\text{eff}(r)=\frac{\mathrm{L}^2}{r^2}\,\left(1-\frac{2\,M\,r^2}{\left(r^2+g^2\right)^{3 / 2}}+\frac{g^2\,r^2}{\left(r^2+g^2\right)^2}+\frac{r^2}{\ell^2_{p}}-c\,r\right).\label{pp6}
\end{eqnarray}

 In Figure \ref{fig:null-potential}, we present the effective potential for null geodesics, varying the ABG NLED parameter $g$, the QF constant $c$, and their combinations. In panel (a) of Figure \ref{fig:null-potential},  we observe that as the value of $g$ increases, the effective potential also increases. Conversely, in panel (b), the effective potential decreases as $c$ increases. However, in panel (c), we find that the effective potential increases when both $g$ and $c$ are increased simultaneously.

In Figure \ref{fig:null-comparison}, we present a comparison of the effective potential for null geodesics across three different BH metrics, highlighting how the potential changes in the presence of various parameters compared to the case without these parameters.

For circular null geodesics, we have the conditions $\dot{r}=0$ and $\ddot{r}=0$ which implies the following relations
\begin{equation}
    V_\text{eff}(r)=\mathrm{E}^2,\quad\quad V'_\text{eff}(r)=0.\label{pp7}
\end{equation}

The first relation gives us the critical impact parameter for photon ray given by
\begin{equation}
    \frac{1}{\beta_c}=\frac{\mathrm{E}}{\mathrm{L}}=\frac{1}{r_c}\,\left(1-\frac{2\,M\,r^2_c}{\left(r^2_c+g^2\right)^{3 / 2}}+\frac{g^2\,r^2_c}{\left(r^2_c+g^2\right)^2}+\frac{r^2_c}{\ell^2_{p}}-c\,r_c\right)^{1/2}.\label{pp8}
\end{equation}
The second relation gives us the photon sphere radius $r=r_c$ given by
\begin{equation}
    2\,f(r_c)=r_c\,f'(r_c).\label{pp9}
\end{equation}

Now, we determine force on photon light and analyze how the the ABG NLED parameter $g$, the radius of curvature $\ell_p$, and the QF parameter $c$ for the chosen state parameter $w=2/3$ influences on it. This force is expressible in terms of the effective potential as 
\begin{equation}
\mathrm{F}_\text{ph}=-\frac{1}{2}\,V'_\text{eff}(r).\label{force}    
\end{equation}

Thereby, using the expression (\ref{pp6}), we find the force as
\begin{eqnarray}
    \mathrm{F}_\text{ph}=\frac{\mathrm{L}^2}{r^3}\,\left(2\,f(r)-r\,f'(r)\right)=\frac{\mathrm{L}^2}{r^3}\,\left[1-\frac{3\,M\,r^4}{(r^2+g^2)^{5/2}}+\frac{2\,g^2\,r^4}{(r^2+g^2)^3}-\frac{c\,r}{2}\right].\label{pp10}
\end{eqnarray}
Equation (\ref{pp10}) is the expression of force on photon particle under the influence of the gravitational field produced by the selected BH. From the expression, it is evident that the force on massless photon particle is influenced by the ABG NLED parameter $g$, and the QF parameter $c$ for the chosen state parameter $w=2/3$ including the BH mass $M$. 

\begin{figure}[ht!]
    \centering
    \subfloat[$c=0.01$]{\centering{}\includegraphics[width=0.4\linewidth]{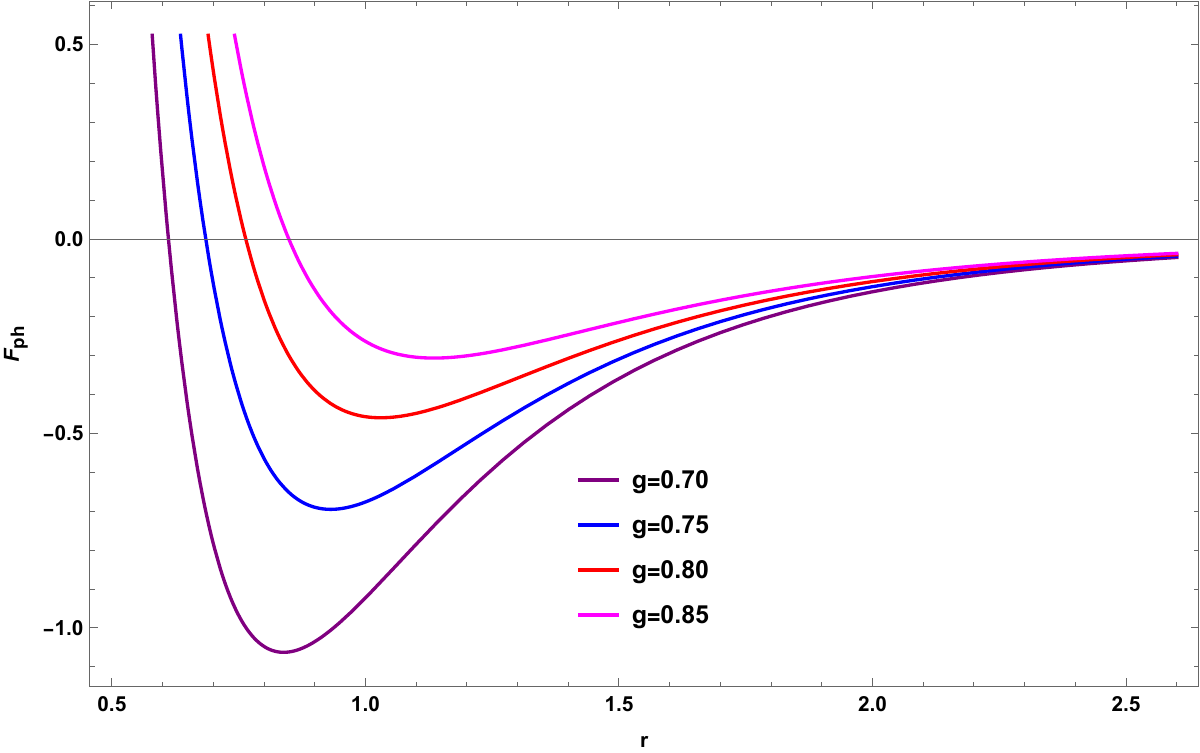}}\quad\quad\quad
    \subfloat[]{\centering{}\includegraphics[width=0.4\linewidth]{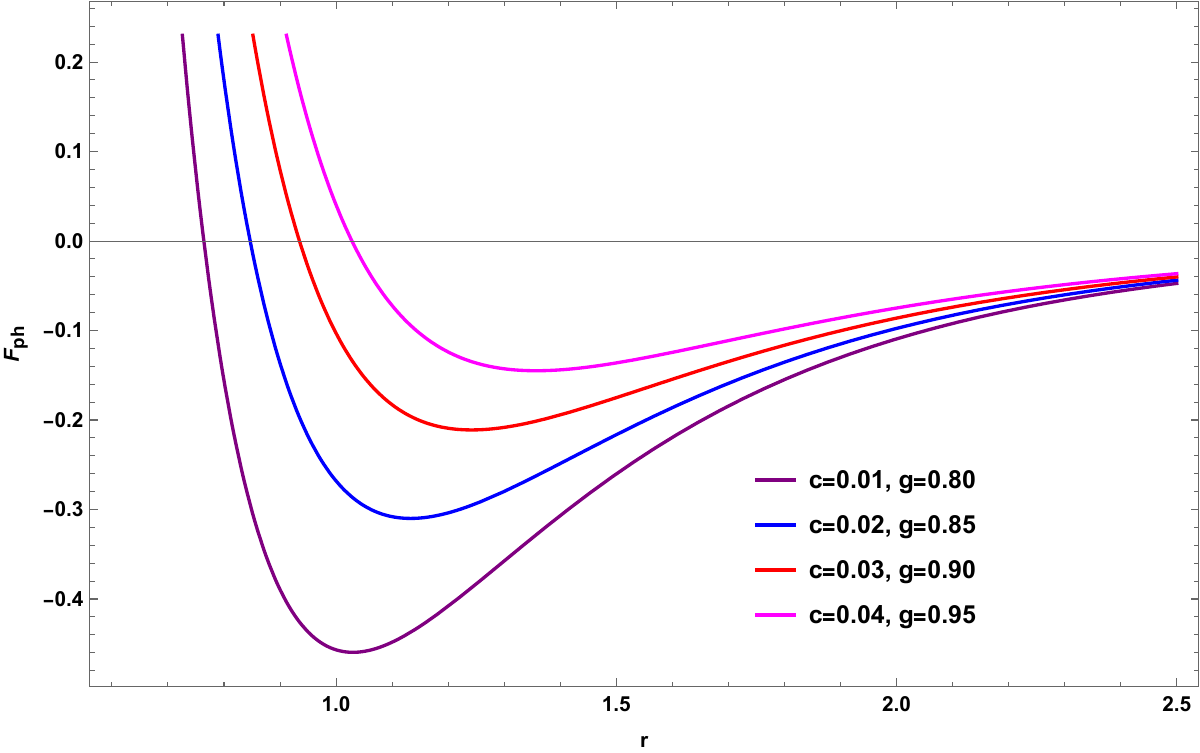}}
    \caption{The behavior of force ($\mathrm{F}_\text{ph}$) on photon particle for different values of $g$ and $c$. Here, we set $M=2$ and $\mathrm{L}=1$.}
    \label{fig:force}
\end{figure}

 In Figure \ref{fig:force}, we illustrate the force on a massless photon particle under the influence of the gravitational field generated by the selected BH, while varying the ABG NLED parameter $g$, the QF constant $c$, and/or their combinations. In panel (a) of Figure \ref{fig:force}, we observe that as the value of $g$ increases, the force also increases. A similar trend is seen in panel (b), where both $g$ and $c$ are increased simultaneously. This suggests that higher values of $g$ and $c$ produce a stronger gravitational field, which influences the photon’s path as it travels in the vicinity of the black hole.

Now, we determine period of circular orbits encompasses the time required for a particle to complete one full revolution around the circular time-like path. The formulas for calculating the period of a circular orbit in proper ($T_{\tau}$) and coordinate times ($T_t$) are derived in Ref. \cite{SF1}. In our case, following the similar approach, we find these times are
\begin{eqnarray}
    T_{\tau}=\frac{2\,\pi\,r^2_c}{\mathrm{L}},\label{pp11}
\end{eqnarray}
where $\mathrm{L}$ is the conserved angular momentum. And
\begin{equation}
    T_t=2\,\pi\,|\beta_c|=\frac{2\,\pi\,r_c}{\sqrt{f(r_c)}}=\frac{2\,\pi\,r_c}{\sqrt{1-\frac{2\,M\,r^2_c}{\left(r^2_c+g^2\right)^{3 / 2}}+\frac{g^2\,r^2_c}{\left(r^2_c+g^2\right)^2}+\frac{r^2_c}{\ell^2_{p}}-c\,r_c}},\label{pp12}
\end{equation}
where $r_c$ can be determined from Eq. (\ref{pp9}).

Now, we focus on an important physical quantity which determine whether the circular orbit is table or unstable. This is determined by the Lyapunov exponent defined as
\begin{equation}
    \lambda^\text{null}_L=\sqrt{-\frac{V''_\text{eff}}{2\,\dot{t}^2}}.\label{pp13}
\end{equation}

Using Eq. (\ref{pp2}), the effective potential potential (\ref{pp6}) and finally using the condition (\ref{pp9}), we find the following relation
\begin{equation}
    \lambda^\text{null}_L=\sqrt{f(r_c)\,\left(\frac{f(r_c)}{r^2_c}-\frac{f''(r_c)}{2}\right)}.\label{pp14}
\end{equation}

Using the metric function (\ref{m1}), we find the following expression:
\begin{eqnarray}
    \lambda^\text{null}_L&=&\frac{1}{r}\,\sqrt{1-\frac{2\,M\,r^2}{\left(r^2+g^2\right)^{3 / 2}}+\frac{g^2\,r^2}{\left(r^2+g^2\right)^2}+\frac{r^2}{\ell^2_{p}}-c\,r}\times\nonumber\\
    &&\sqrt{1-c\,r-\frac{12\,g^2\,r^6}{(g^2 + r^2)^4} +\frac{15\, M\,r^6}{(g^2 + r^2)^{7/2}}+\frac{10\, g^2\, r^4}{(g^2 + r^2)^3}-\frac{15\, M\, r^4}{(g^2 + r^2)^{5/2}}}.\label{pp15}
\end{eqnarray}

From the above equation (\ref{pp15}), it is evident that several factors such as ABG NLED parameter $g$, the radius of curvature $\ell_p$, and the QF parameter $c$ for the chosen specific state parameter $w=2/3$ influences the Lyapunov exponent.

\begin{figure}[ht!]
    \centering
    \subfloat[$c=0.01$]{\centering{}\includegraphics[width=0.4\linewidth]{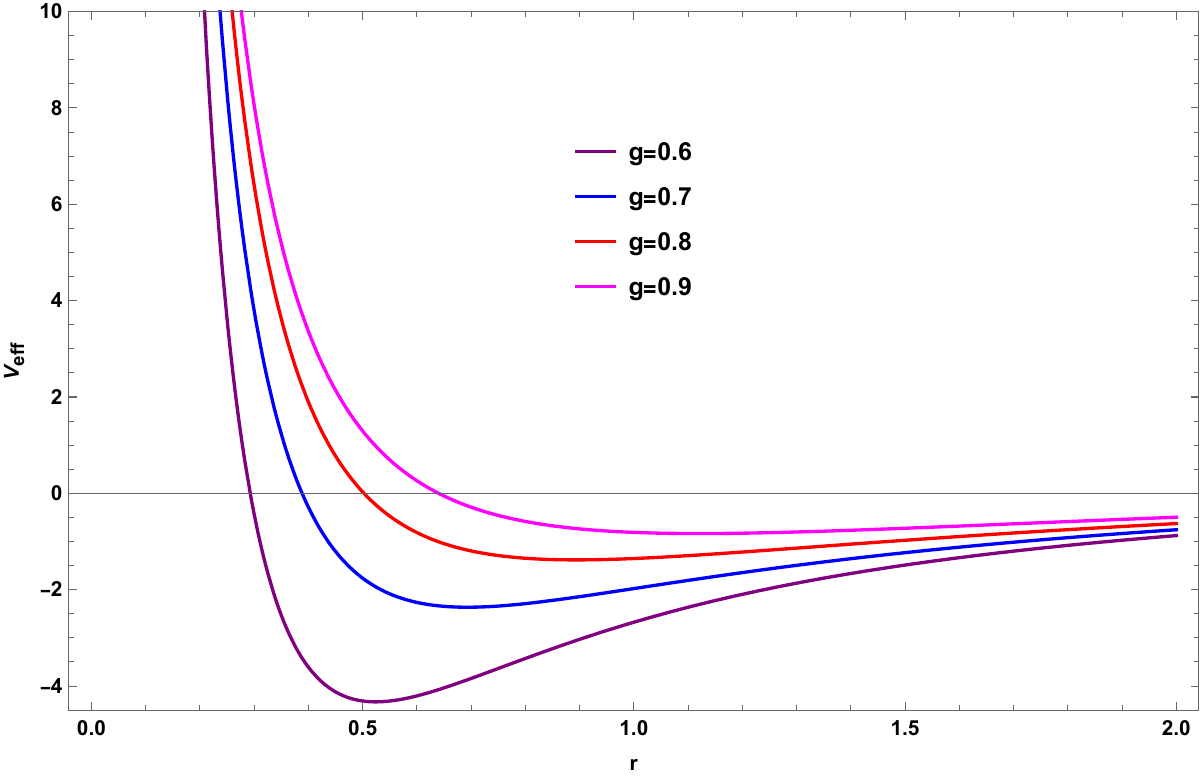}}\quad\quad\quad
    \subfloat[$g=0.9$]{\centering{}\includegraphics[width=0.4\linewidth]{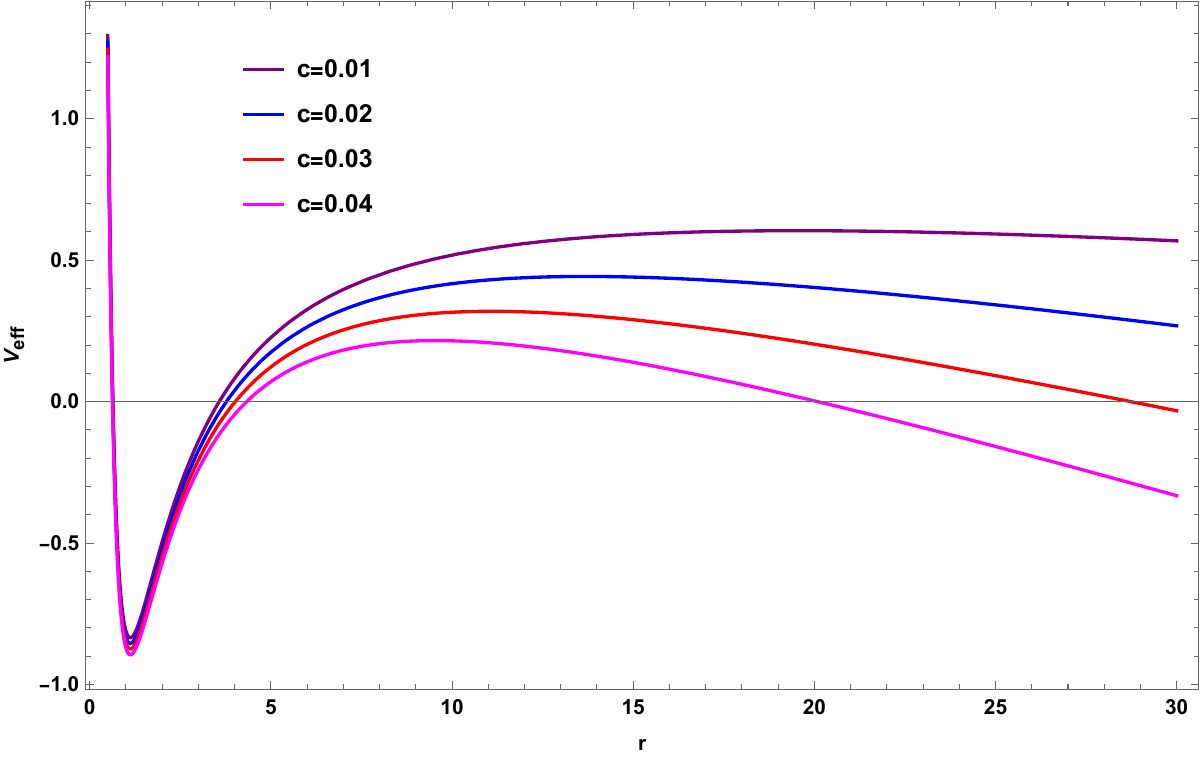}}\\
    \subfloat[]{\centering{}\includegraphics[width=0.4\linewidth]{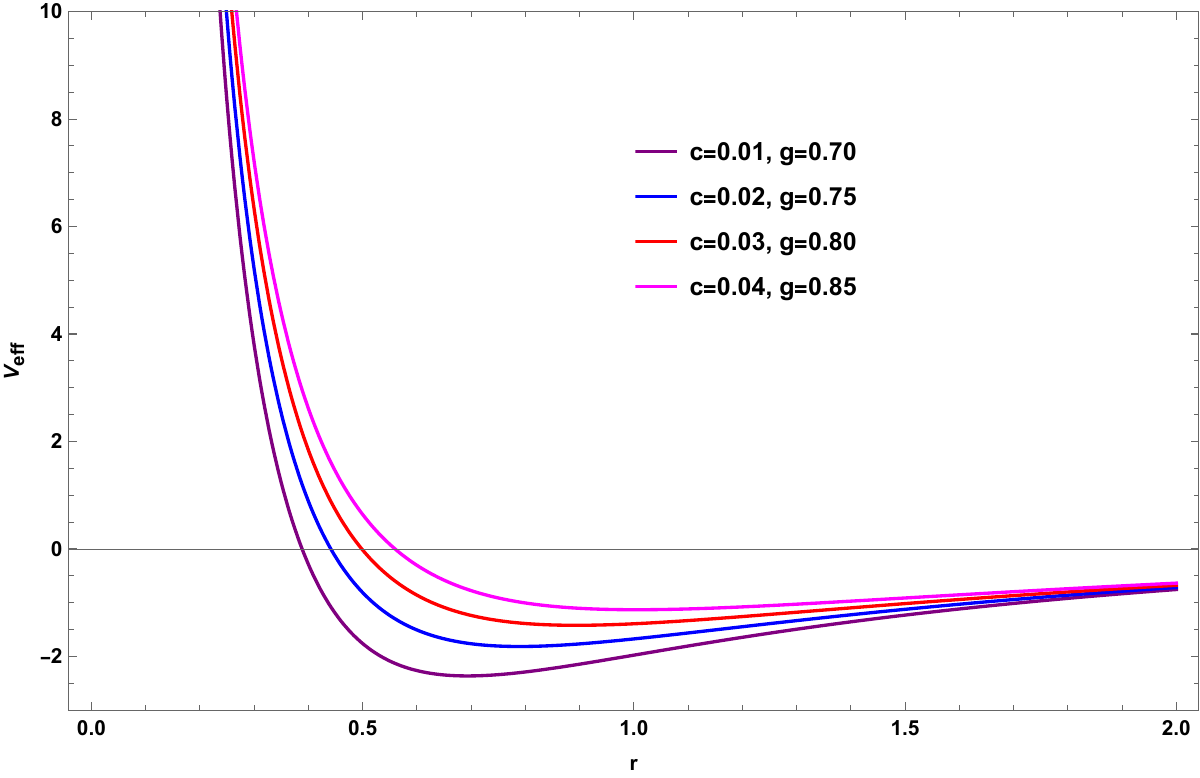}}
    \caption{The behavior of the effective potential $V_\text{eff}$ for time-like geodesics for different values of $g$ and $c$. Here, we set $M=2$ and $\mathrm{L}=1$.}
    \label{fig:timelike-potential}
\end{figure}

\subsection{Massive test particles}

In this part, we study motion of massive test particles and discuss how the ABG NLED parameter $g$, the radius of curvature $\ell_p$, and the QF parameter $c$ for the chosen state parameter $w=2/3$ influences time-like particles. For time-like geodesics, $\kappa=1-$, the effective potential from Eq. (\ref{pp5}) becomes
\begin{eqnarray}
    V_\text{eff}(r)=\left(1+\frac{\mathrm{L}^2}{r^2}\right)\,\left(1-\frac{2\,M\,r^2}{\left(r^2+g^2\right)^{3 / 2}}+\frac{g^2\,r^2}{\left(r^2+g^2\right)^2}+\frac{r^2}{\ell^2_{p}}-c\,r\right).\label{qq1}
\end{eqnarray}

 In Figure \ref{fig:timelike-potential}, we present the effective potential for time-like geodesics, varying the ABG NLED parameter $g$, the QF constant $c$, and their combinations. In panel (a) of Figure \ref{fig:timelike-potential},  we observe that as the value of $g$ increases, the effective potential also increases. In contrast, in panel (b), the effective potential decreases as $c$ increases. However, in panel (c), we find that the effective potential increases when both $g$ and $c$ are increased simultaneously.

For circular motions of time-like particles in the equatorial plane, we have the conditions 
\begin{equation}
    \dot{r}=0\Rightarrow V_\text{eff}(r)=\mathrm{E}^2,\quad\quad  \ddot{r}=0\Rightarrow V'_\text{eff}(r)=0,\label{qq2}
\end{equation}
where prime denotes partial derivative w. r. t. $r$.

Using Eq. (\ref{qq1}) into above the relation (\ref{qq2}) gives us the following physical quantities:
\begin{eqnarray}
    &&\mathrm{L}=r\,\sqrt{\frac{-\frac{c\,r}{2}-\frac{2\,g^2\,r^4}{(g^2 + r^2)^3}+\frac{3\,M\,r^4}{(g^2 + r^2)^{5/2}}+\frac{g^2\,r^2}{(g^2+r^2)^2}-\frac{2\,M\,r^2}{(g^2+r^2)^{3/2}}+\frac{r^2}{\ell^2_p}}{1-\frac{c\,r}{2}+\frac{2\,g^2\,r^4}{(g^2 + r^2)^3}-\frac{3\,M\,r^4}{(g^2+r^2)^{5/2}}}},\label{qq3}\\
    &&\mathrm{E}=\pm\,\frac{\left(1-\frac{2\,M\,r^2}{\left(r^2+g^2\right)^{3 / 2}}+\frac{g^2\,r^2}{\left(r^2+g^2\right)^2}+\frac{r^2}{\ell^2_{p}}-c\,r\right)}{\sqrt{1-\frac{c\,r}{2}+\frac{2\,g^2\,r^4}{(g^2 + r^2)^3}-\frac{3\,M\,r^4}{(g^2+r^2)^{5/2}}}},\label{qq4}
\end{eqnarray}

The above quantities are respectively, the energy and the angular momentum associated with a time-like particle orbiting in circular geodesics in the equatorial plane. From the equations (\ref{qq3})--(\ref{qq4}), it is evident that various factors such as ABG NLED parameter $g$, the radius of curvature $\ell_p$, and the QF parameter $c$ for the chosen specific state parameter $w=2/3$ influences the these physical quantities $(\mathrm{E}\,,\,\mathrm{L})$. Moreover, the BH mass $M$ also influence these quantities of time-like particle.  

\begin{figure}[ht!]
    \centering
    \subfloat[$c=0.1$]{\centering{}\includegraphics[width=0.4\linewidth]{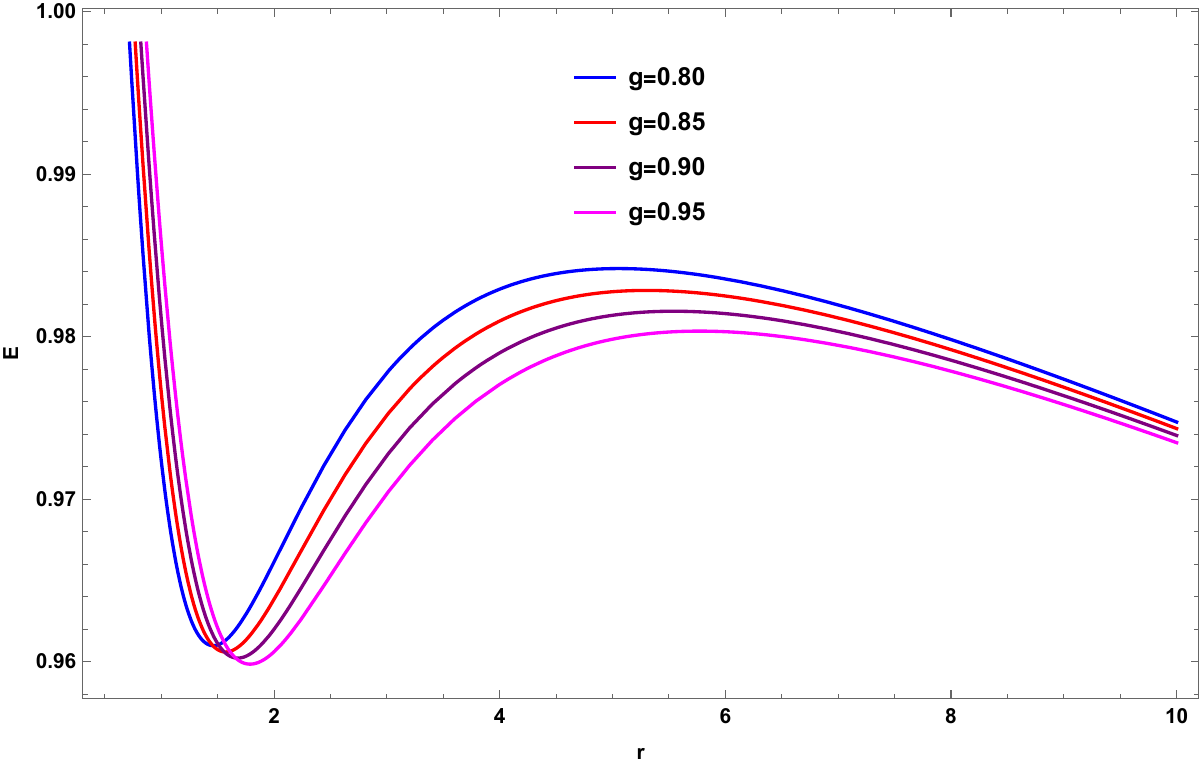}}\quad\quad\quad
    \subfloat[$g=0.3$]{\centering{}\includegraphics[width=0.4\linewidth]{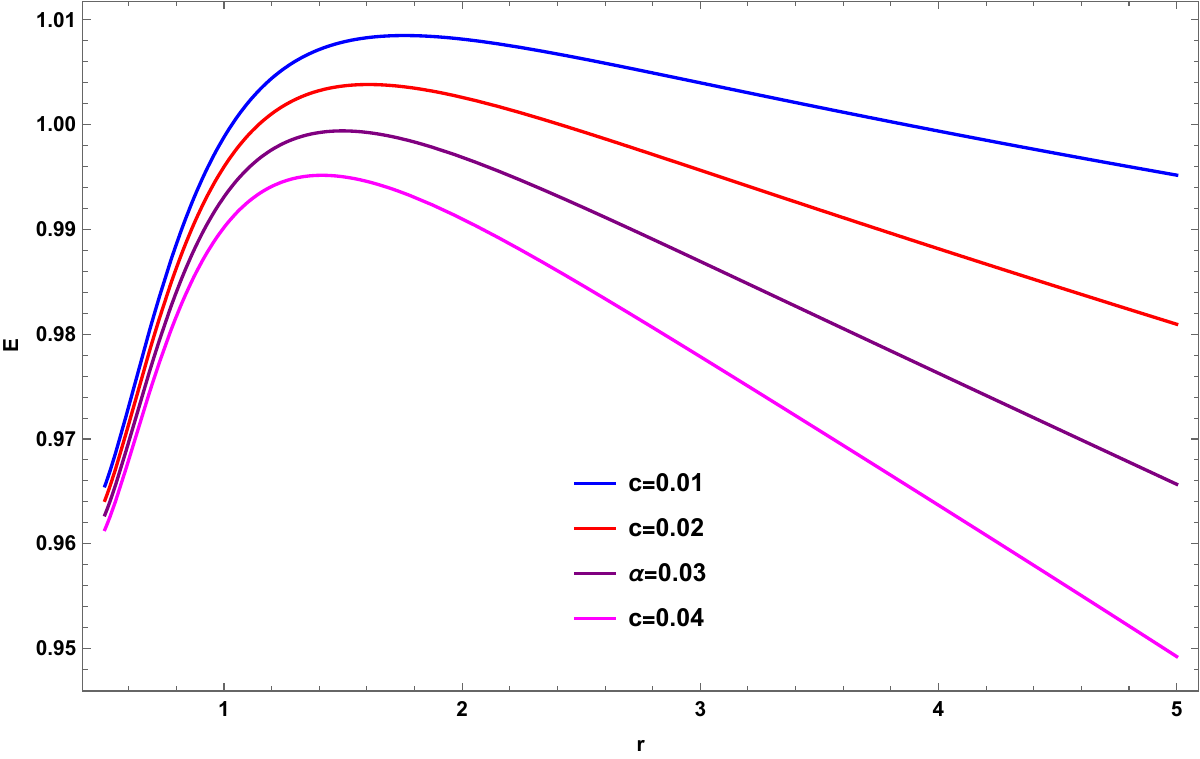}}\\
    \subfloat[]{\centering{}\includegraphics[width=0.4\linewidth]{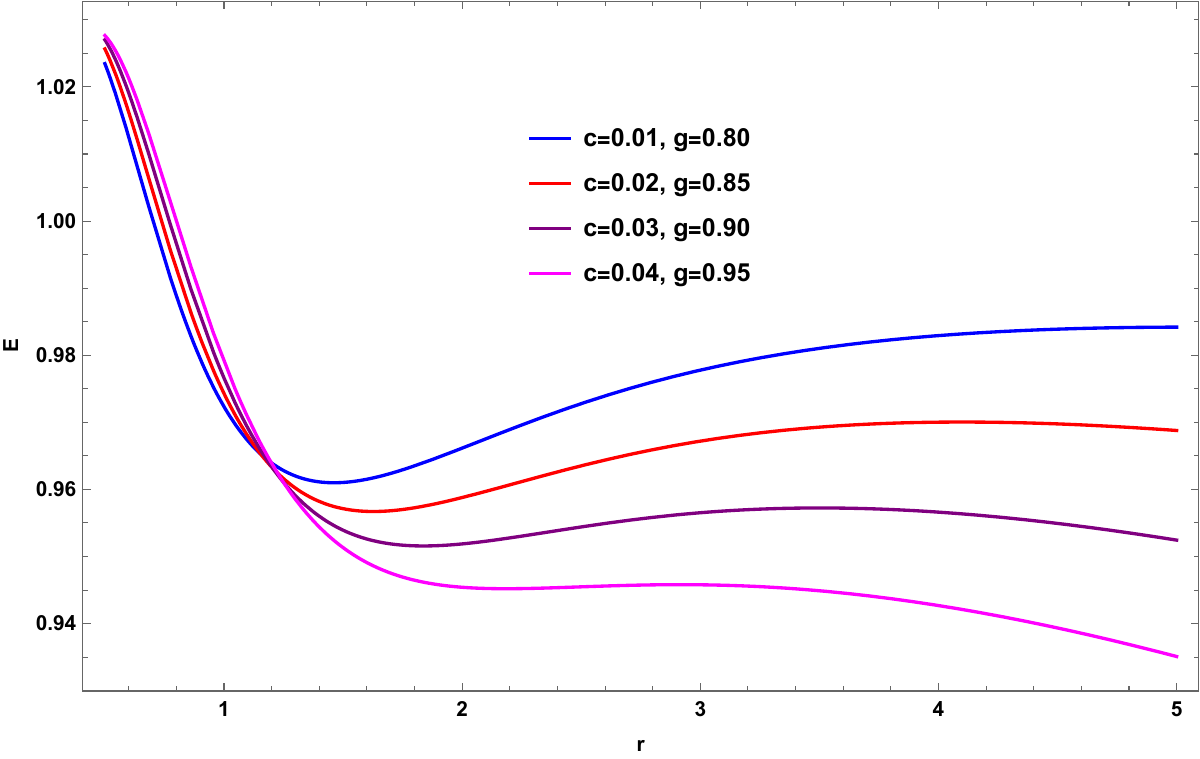}}
    \caption{The behavior of time-like particles' energy $\mathrm{E}$ for different values of $g$ and $c$. Here, we set $M=0.1$ and $\ell_p=10$.}
    \label{fig:energy}
\end{figure}

 In Figure \ref{fig:energy}, we depict the energy of the time-like particles, varying the ABG NLED parameter $g$, the QF constant $c$, and their combinations. In all panels (a)--(c) of Figure \ref{fig:energy}, we observe that as the values of $g$, $c$, and their combinations ($g, c$) increase, the energy decreases. This suggests that higher values of $g$ and $c$ generate a stronger gravitational field, which affects the motion of time-like particles on circular orbits in the equatorial plane near the black hole, leading to a reduction in particles' energy.

Now, we define orbital velocity of test particles on circular orbits as follows:
\begin{equation}
    \Omega \equiv\frac{\dot{\phi}}{\dot{r}}=\sqrt{\frac{f'(r)}{2\,r}}=\sqrt{-\frac{c}{2\,r}-\frac{2\,g^2\,r^2}{(g^2 + r^2)^3}+\frac{3\,M\,r^2}{(g^2 + r^2)^{5/2}}+\frac{g^2}{(g^2+r^2)^2}-\frac{2\,M}{(g^2+r^2)^{3/2}}+\frac{1}{\ell^2_p}}.\label{qq5}
\end{equation}
And the proper angular velocity is defined by $\omega=\dot{\phi}=\mathrm{L}/r^2$, and so we have that
\begin{equation}
    \omega=\sqrt{\frac{2}{2\,f-r\,f'}}\,\Omega=\frac{\Omega}{\sqrt{1-\frac{c\,r}{2}+\frac{2\,g^2\,r^4}{(g^2 + r^2)^3}-\frac{3\,M\,r^4}{(g^2+r^2)^{5/2}}}}.\label{qq6}
\end{equation}

To determine the geodesic precession frequency $\Theta_{GPF}$ for a gyroscope within the GP-B, we use Eq. (\ref{qq6}) and following the methodology in Ref. \cite{RS1}, we obtain  
\begin{equation}
    \Theta_{GPF}=\Omega-\Omega_{GPF},\quad\quad \Omega_{GPF}=\Omega\,\sqrt{f-\Omega^2\,r^2}.\label{qq7}
\end{equation}

Substituting Eqs. (\ref{qq5}) and (\ref{qq6}) into Eq. (\ref{qq7}), we find
\begin{eqnarray}
    \Theta_{GPF}&=&\sqrt{\frac{f'(r)}{2\,r}}\,\left[1-\sqrt{\frac{1}{2}\,\left(2\,f-r\,f'\right)}\right]\nonumber\\
    &=&\sqrt{-\frac{c}{2\,r}-\frac{2\,g^2\,r^2}{(g^2 + r^2)^3}+\frac{3\,M\,r^2}{(g^2 + r^2)^{5/2}}+\frac{g^2}{(g^2+r^2)^2}-\frac{2\,M}{(g^2+r^2)^{3/2}}+\frac{1}{\ell^2_p}}\times\nonumber\\
    &&\left[1-\sqrt{1-\frac{c\,r}{2}+\frac{2\,g^2\,r^4}{(g^2 + r^2)^3}-\frac{3\,M\,r^4}{(g^2+r^2)^{5/2}}}\right].\label{qq8}
\end{eqnarray}

From the above expression (\ref{qq8}), it is evident that various factors such as ABG NLED parameter $g$, the radius of curvature $\ell_p$, and the QF parameter $c$ for the chosen specific state parameter $w=2/3$ influences the geodesic precession frequency. Moreover, the mass $M$ also influence this frequency. 

\begin{figure}[ht!]
    \centering
    \subfloat[$c=0.01$]{\centering{}\includegraphics[width=0.4\linewidth]{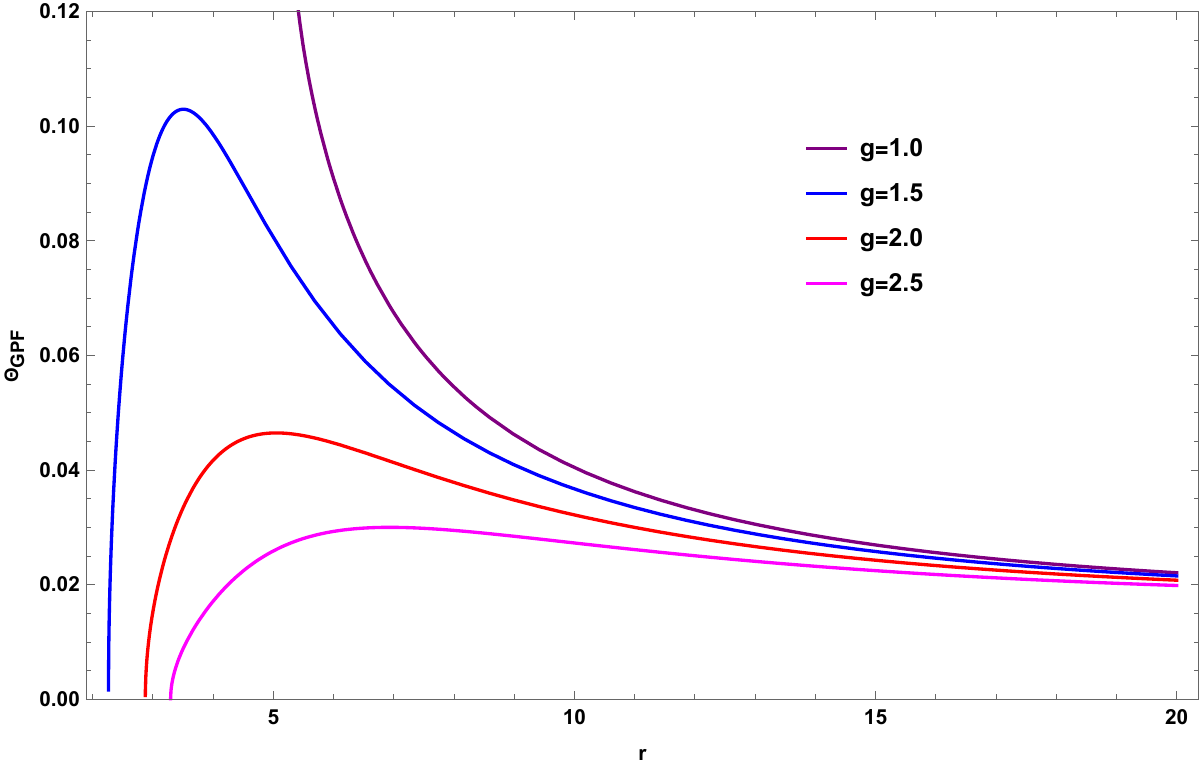}}\quad\quad\quad
    \subfloat[$g=0.5$]{\centering{}\includegraphics[width=0.4\linewidth]{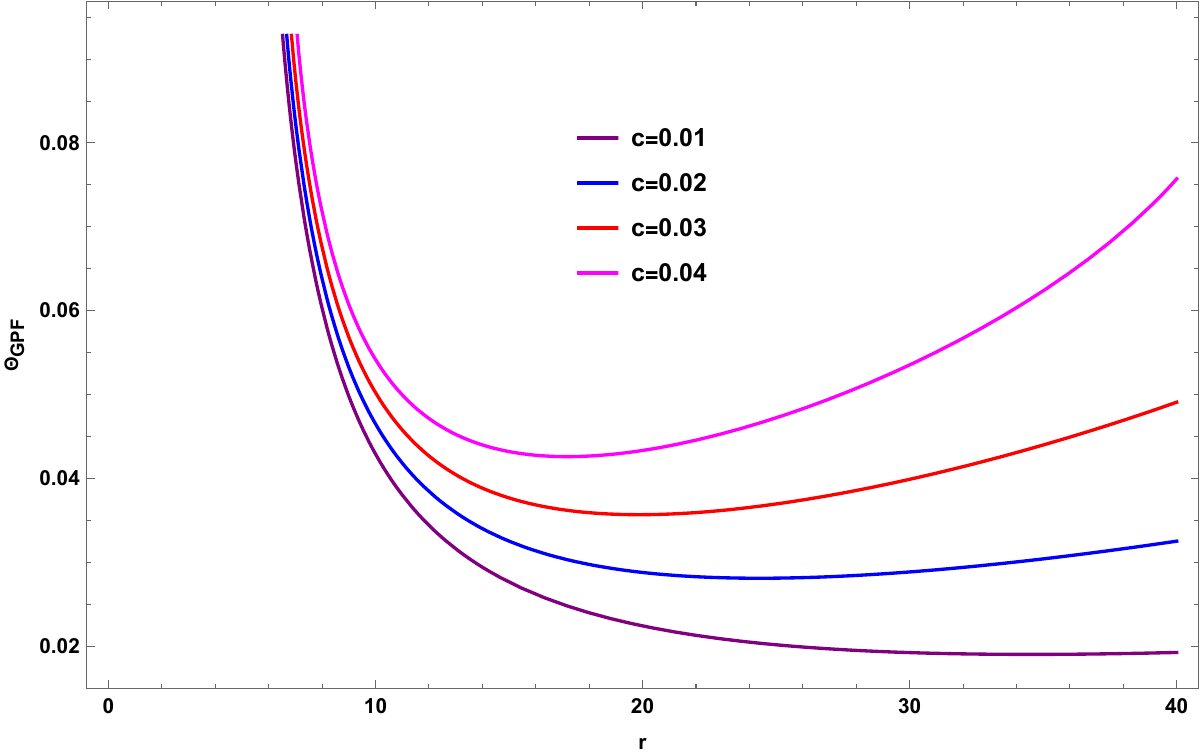}}\\
    \subfloat[]{\centering{}\includegraphics[width=0.4\linewidth]{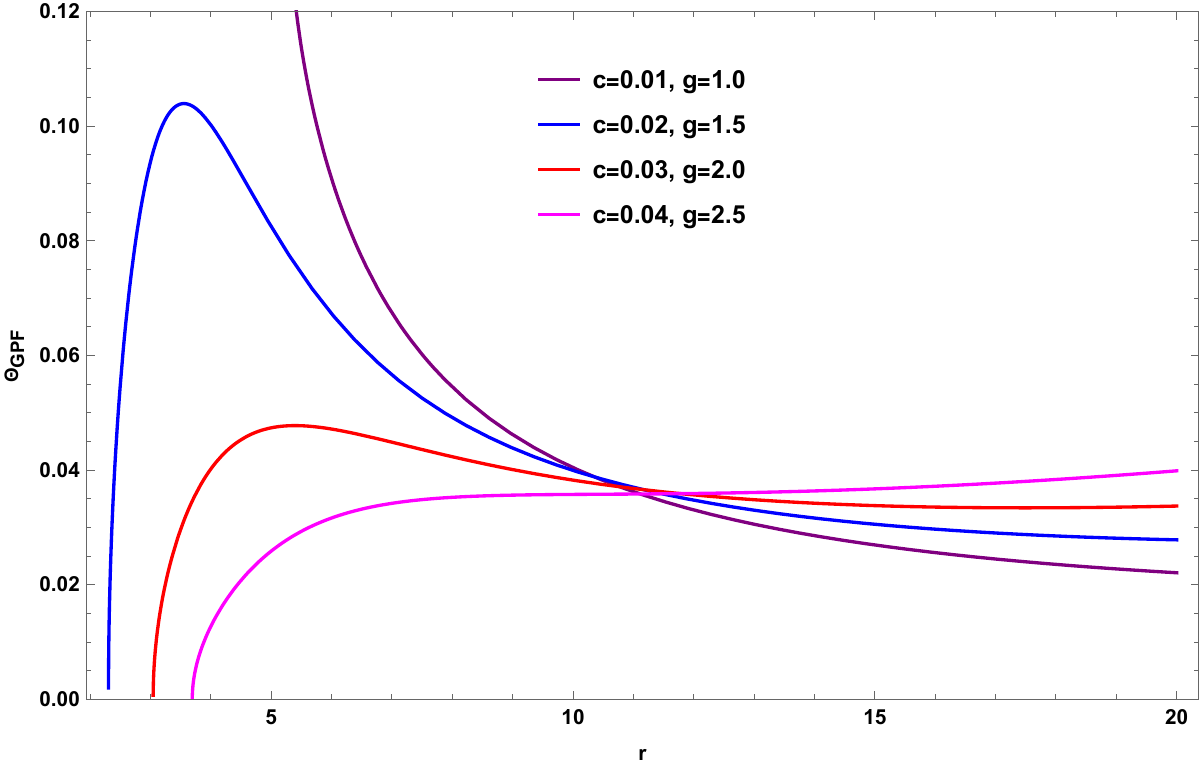}}
    \caption{The behavior of the geodesic precession frequency $\Theta_\text{GPF}$ for different values of $g$ and $c$. Here, we set $M=2$ and $\ell_p=10$.}
    \label{fig:frequency}
\end{figure}

\begin{figure}[ht!]
    \centering
    \includegraphics[width=0.4\linewidth]{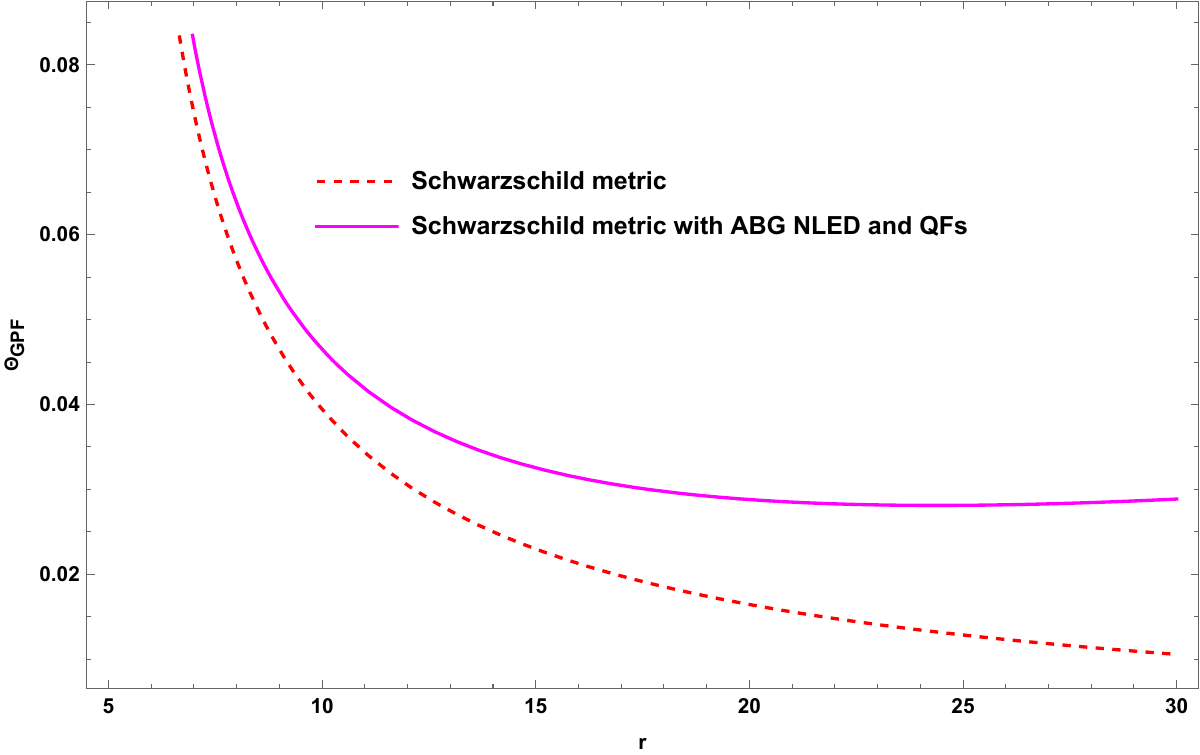}
    \caption{A comparison of the geodesic precession frequency $\Theta_\text{GPF}$ for different BHs. Here, we set $g=0.5$, $c=0.02$, $M=2$ and $\ell_p=10$.}
    \label{fig:frequency-comp}
\end{figure}

 In Figure \ref{fig:frequency}, we illustrate the precession frequency for time-like particles, considering variations in the ABG NLED parameter $g$, the QF constant $c$, and their combinations. In panel (a), we observe that as the value of $g$ increases, the precession frequency decreases. Conversely, in panel (b), the frequency increases with an increase in $c$. In panel (c), however, we find that the precession frequency decreases when both $g$ and $c$ are increased simultaneously. This suggests that, for a fixed QF parameter $c$ and a specific state parameter $w = -2/3$, higher values of the ABG parameter $g$ have a more pronounced effect on the motion of time-like particles in circular orbits within the equatorial plane.

In Figure \ref{fig:frequency-comp}, we present a comparison of the precession frequency of time-like geodesics across two different BH metrics, highlighting how the frequency changes in the presence of parameters ($g, c$) compared to the case without these parameters.

\begin{figure}[ht!]
    \centering
    \subfloat[$c=0.01$]{\centering{}\includegraphics[width=0.4\linewidth]{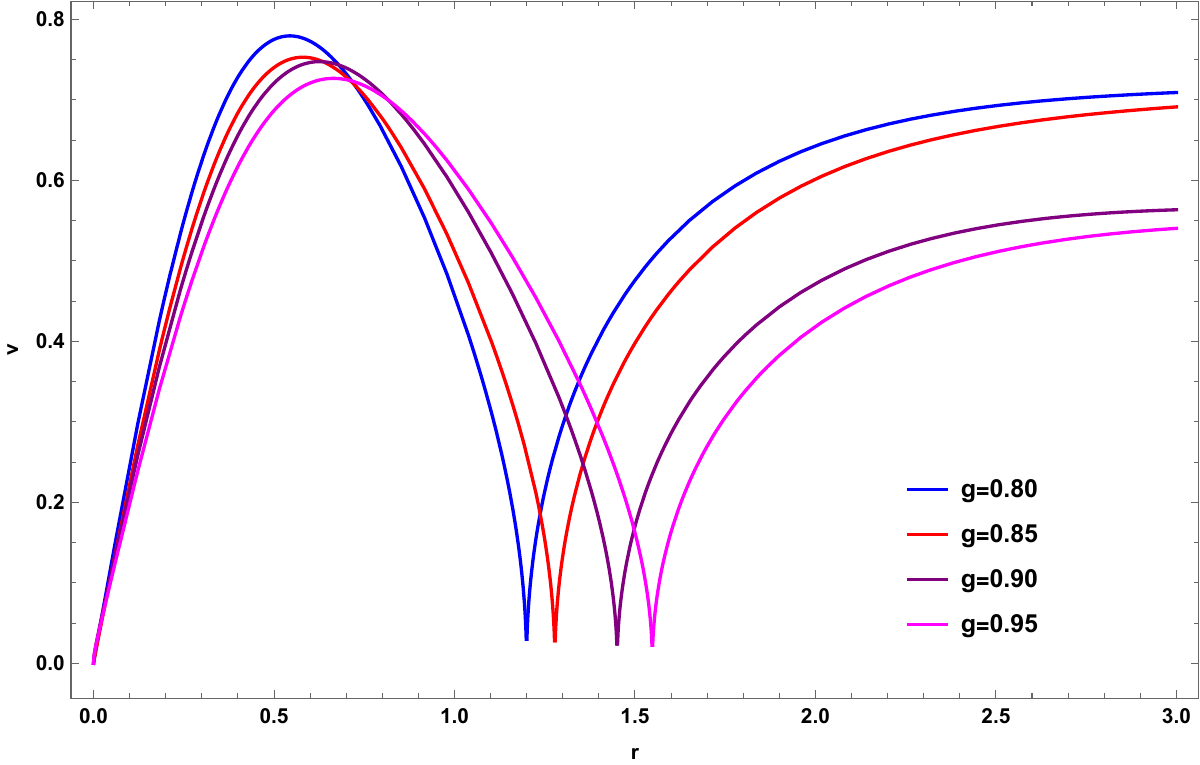}}\quad\quad\quad
    \subfloat[$g=0.5$]{\centering{}\includegraphics[width=0.4\linewidth]{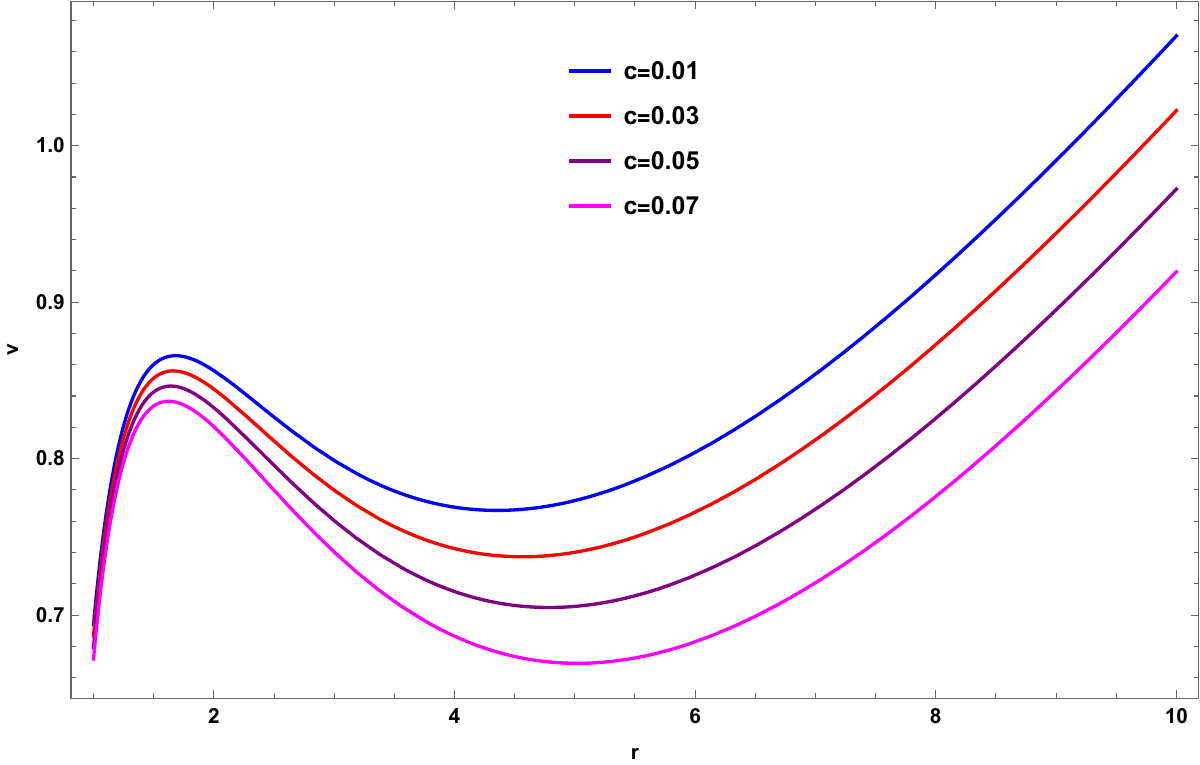}}\\
    \subfloat[]{\centering{}\includegraphics[width=0.4\linewidth]{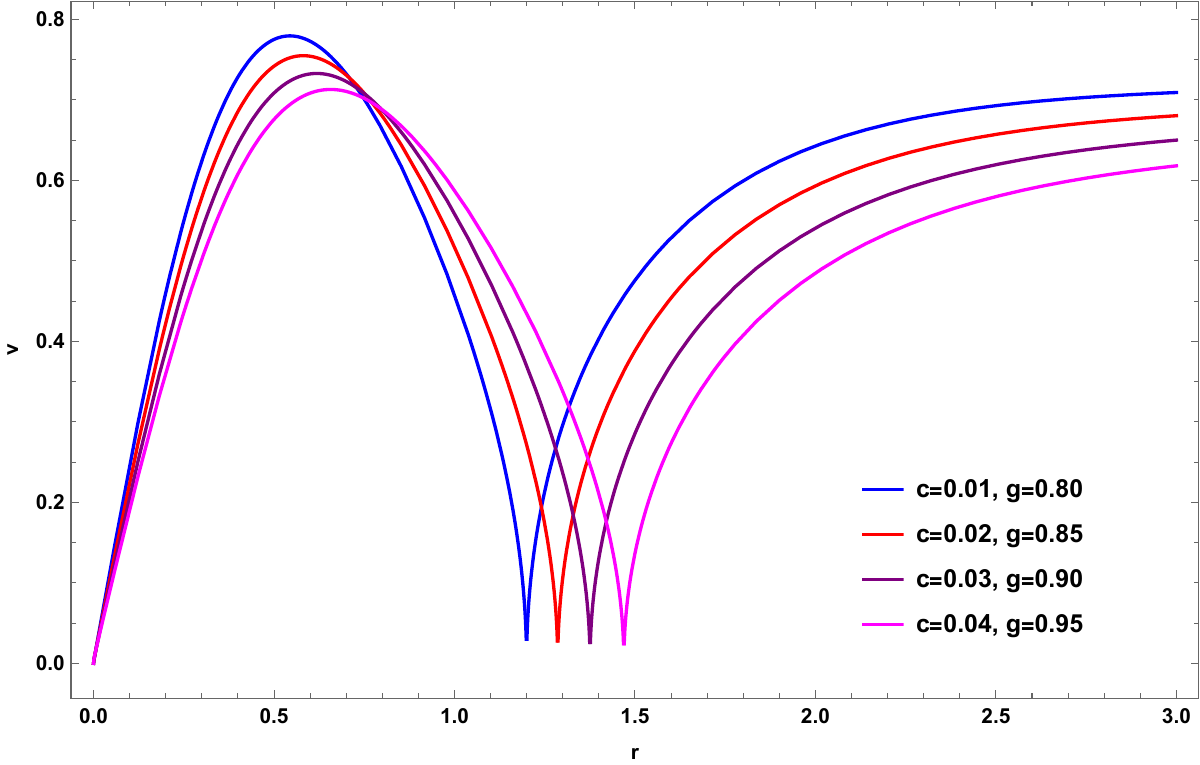}}
    \caption{The behavior of the circular speed $v$ of time-like particle for for different values of $g$ and $c$. Here, we set $M=2$ and $\ell_p=10$.}
    \label{fig:speed}
\end{figure}

\begin{figure}[ht!]
    \centering
    \includegraphics[width=0.4\linewidth]{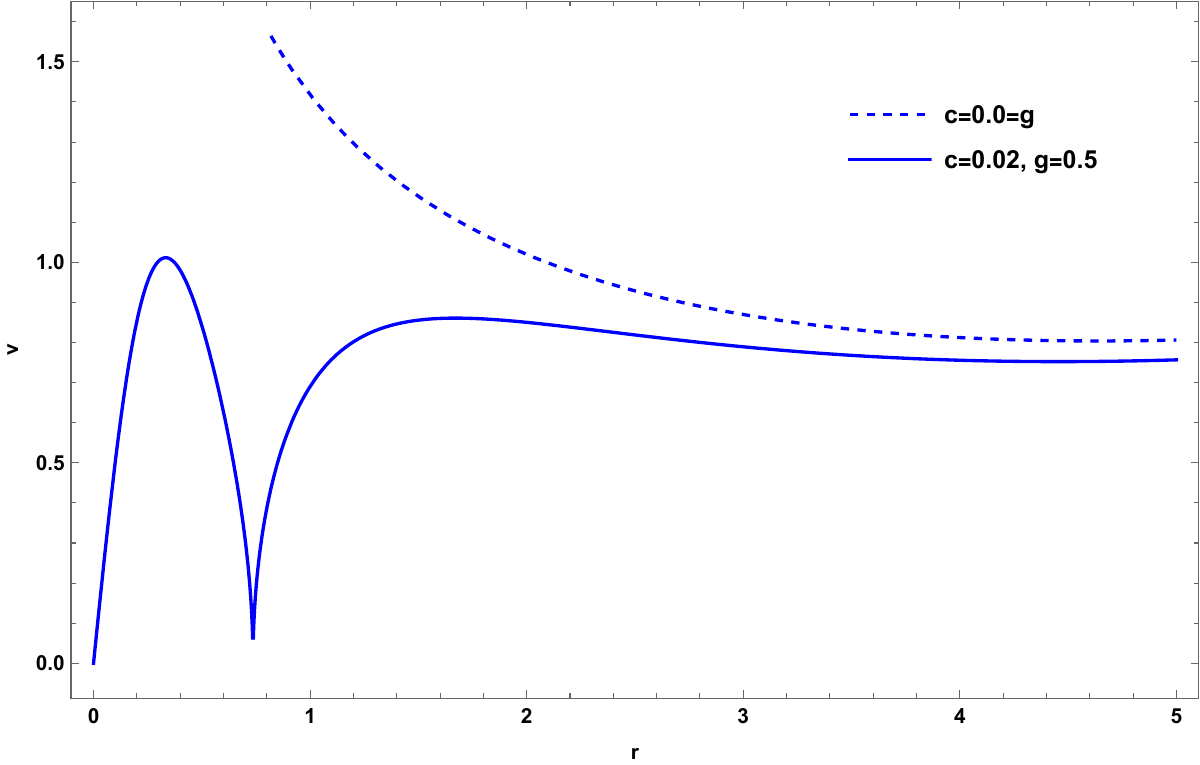}
    \caption{A comparison of the circular speed $v$ of time-like particle for different BHs. Here, we set $c=0.02$, $g=0.5$, $M=2$ and $\ell_p=10$.}
    \label{fig:speed-comparison}
\end{figure}

Finally, we aim to find speed with which time-like particle orbit the BH at a very large distance in comparison with the horizon of the BH. This is in analogy with a distant star in a galaxy moving in a circle around the BH of the galaxy. In the zeroth approximation, one write
\begin{equation}
    f(r)=1+2\,\Phi(r)\label{qq9}
\end{equation}
in which $\Phi(r)$ is the Newtonian gravitational potential for the time-like particle of unit mass. 

Explicitly, we obtain the the gravitational potential $\Phi(r)$ using the metric function (\ref{m1}) as:
\begin{equation}
    \Phi(r)=\frac{1}{2}\,\left(-\frac{2\,M\,r^2}{\left(r^2+g^2\right)^{3 / 2}}+\frac{g^2\,r^2}{\left(r^2+g^2\right)^2}+\frac{r^2}{\ell^2_{p}}-c\,r\right).\label{qq10}
\end{equation}

The effective gravitational force is defined by the following formula 
\begin{equation}
    \mathrm{F}_c=-\frac{\partial \Phi(r)}{\partial r},\label{qq11}    
\end{equation}

Using potential (\ref{qq10}), we find
\begin{equation}
    \mathrm{F}_c=\frac{c}{2}+\frac{2\,g^2\,r^3}{(g^2 + r^2)^3}-\frac{3\,M\,r^3}{(g^2 + r^2)^{5/2}}-\frac{g^2\,r}{(g^2+r^2)^2}+\frac{2\,M\,r}{(g^2+r^2)^{3/2}}-\frac{r}{\ell^2_p},\label{qq12}    
\end{equation}
This central force $\mathrm{F}_c$ can be equated with centripetal acceleration, {\it i.e.}, $|\mathrm{F}_c|=\frac{v^2}{r}$ in which $v$ is the speed of time-like particle in circular orbit. This equation results an expression for the circular speed $v$ given by
\begin{equation}
    v=\sqrt{r\,|\mathrm{F}_c|}=\sqrt{\left|\frac{c\,r}{2}+\frac{2\,g^2\,r^4}{(g^2 + r^2)^3}-\frac{3\,M\,r^4}{(g^2 + r^2)^{5/2}}-\frac{g^2\,r^2}{(g^2+r^2)^2}+\frac{2\,M\,r^2}{(g^2+r^2)^{3/2}}-\frac{r^2}{\ell^2_p}\right|}.\label{qq13}
\end{equation}

The expression (\ref{qq13}) shows that the circular speed of time-like particles in orbiting is influenced by ABG NLED parameter $g$, the radius of curvature $\ell_p$, and the QF parameter $c$ for the chosen specific state parameter $w=2/3$ including the BH mass $M$. 

 In Figure \ref{fig:speed}, we depict the angular speed $v$ of the time-like particles, varying the ABG NLED parameter $g$, the QF constant $c$, and their combinations. In all panels (a)--(c) of Figure \ref{fig:speed}, we observe that as the values of $g$, $c$, and their combinations ($g, c$) increase, the speed decreases. This suggests that higher values of $g$ and $c$ may generate a stronger gravitational field, which affects the motion of time-like particles on circular orbits in the equatorial plane around the black hole, leading to a reduction in particles' speed.

In Figure \ref{fig:speed-comparison}, we present a comparison of the speed of time-like particle across two different BH metrics, highlighting how the speed changes in the presence of parameters $g \neq 0,\, c \neq 0$ compared to the case without these parameters, that is, the standard Schwarzschild BH.

\section{Scalar Perturbations: The Massless Scalar Field}\label{sec:4}

In this section, we investigate the dynamics of spin-0 massless scalar fields in the background of ABG BHs with QF, focusing on how these perturbations propagate and scatter in the modified spacetime geometry.

The massless scalar field wave equation is governed by the Klein-Gordon equation in curved spacetime:
\begin{equation}
\frac{1}{\sqrt{-g}}\,\partial_{\mu}\left(\sqrt{-g}\,g^{\mu\nu}\,\partial_{\nu}\right)\,\Psi=0,\label{kk1}    
\end{equation}
where $\Psi$ represents the wave function of the scalar field, $g_{\mu\nu}$ is the covariant metric tensor, $g=\det(g_{\mu\nu})$ is the metric determinant, $g^{\mu\nu}$ is the contravariant metric tensor, and $\partial_{\mu}$ denotes the partial derivative with respect to the coordinate system.

To facilitate analytical treatment, we introduce the tortoise coordinate transformation:
\begin{eqnarray}
   dr_*=\frac{dr}{f(r)}\label{kk2}
\end{eqnarray}
which maps the event horizon to $r_* \to -\infty$ while preserving the asymptotic behavior at spatial infinity. Applying this transformation to the line element in Eq. (\ref{metric}) yields:
\begin{equation}
   ds^2=f(r_*)\,\left(-dt^2+dr^2_{*}\right)+h^2(r_*)\,\left(d\theta^2+\sin^2 \theta\,d\phi^2\right),\label{kk3}
\end{equation}
where $f(r_*)$ and $h(r_*)$ are the metric functions expressed in terms of the tortoise coordinate.

Employing the standard separation of variables approach, we adopt the following ansatz for the scalar field:
\begin{equation}
   \Psi(t, r_{*},\theta, \phi)=\exp(i\,\omega\,t)\,Y^{m}_{\ell} (\theta,\phi)\,\frac{\psi(r_*)}{r_{*}},\label{kk4}
\end{equation}
where $\omega$ is the (possibly complex) temporal frequency, $\psi(r)$ is the radial wave function, and $Y^{m}_{\ell} (\theta,\phi)$ represents the spherical harmonics characterized by the angular momentum quantum numbers $\ell$ and $m$.

Substituting this ansatz into the Klein-Gordon equation (\ref{kk1}) and performing the separation of variables, we obtain the radial wave equation in Schr\"{o}dinger-like form:
\begin{equation}
   \frac{\partial^2 \psi(r_*)}{\partial r^2_{*}}+\left(\omega^2-\mathcal{V}\right)\,\psi(r_*)=0,\label{kk5}
\end{equation}
where the effective perturbative potential $\mathcal{V}(r)$ is given by:
\begin{eqnarray}
\mathcal{V}(r)&=&\left(\frac{\ell\,(\ell+1)}{r^2}+\frac{f'(r)}{r}\right)\,f(r)\nonumber\\
&=&\left(\frac{\ell\,(\ell+1)}{r^2}-\frac{c}{r}-\frac{4\,g^2\,r^2}{(g^2 + r^2)^3}+\frac{6\,M\,r^2}{(g^2 + r^2)^{5/2}}+\frac{2\,g^2}{(g^2+r^2)^2}-\frac{4\,M}{(g^2+r^2)^{3/2}}+\frac{2}{\ell^2_p}\right)\times\nonumber\\
&&\left(1-\frac{2\,M\,r^2}{\left(r^2+g^2\right)^{3 / 2}}+\frac{g^2\,r^2}{\left(r^2+g^2\right)^2}+\frac{r^2}{\ell^2_{p}}-c\,r\right).\label{kk6}
\end{eqnarray}

This effective potential (\ref{kk6}) encapsulates the combined effects of the mass, $M$, the angular momentum barrier, $\ell$, the AdS curvature radius, $\ell_p$, the ABG NLED parameter $g$, and QFs parameter $c$ for the chosen specific state parameter $w=2/3$. The intricate mathematical structure of $\mathcal{V}(r)$ reflects the complex gravitational environment encountered by propagating scalar waves, with significant implications for both perturbative stability and radiation characteristics.

In the limit where $g=0$, that is, without the ABG parameter, the scalar perturbative potential from Eq. (\ref{kk6}) becomes
\begin{eqnarray}
\mathcal{V}(r)&=&\left(\frac{\ell\,(\ell+1)}{r^2}-\frac{c}{r}+\frac{2\,M}{r^3}+\frac{2}{\ell^2_p}\right)\,\left(1-\frac{2\,M}{r}+\frac{r^2}{\ell^2_{p}}-c\,r\right).\label{kk7}
\end{eqnarray}
which is similar to the expression obtained for the Kiselev AdS BH metric with a specific state parameter $w=-2/3$.

\section{BH Shadow}\label{sec:5}

Various approaches have been proposed for determining the shadow radius of a spherical BH, as summarized in \cite{shad1}. This involves using the following equation from \cite{shad1} to find the photon sphere radius for a static spherically symmetric metric:
\begin{equation}
    r\,D'(r)=D(r),
\end{equation}
where $D=\sqrt{f}$. 

\begin{figure}[ht!]
    \centering
    \includegraphics[width=0.45\linewidth]{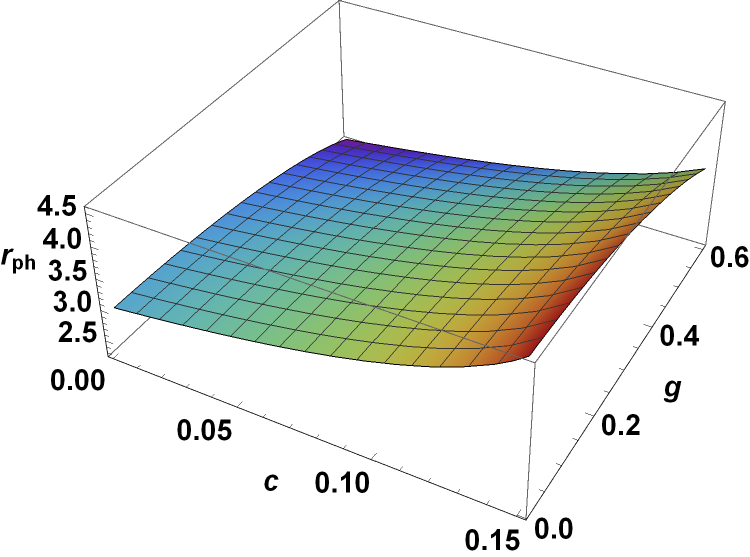}
    \caption{The profile of the photon sphere radius for various values of BH parameters $c$  and $g$ showing that the $r_{ph}$ decreases with $g$ but increases with $c$. Here, $M=1$,  $\ell_p=100$.}
    \label{figa1a}
\end{figure}

\begin{figure}[ht!]
    \centering
    \includegraphics[width=0.45\linewidth]{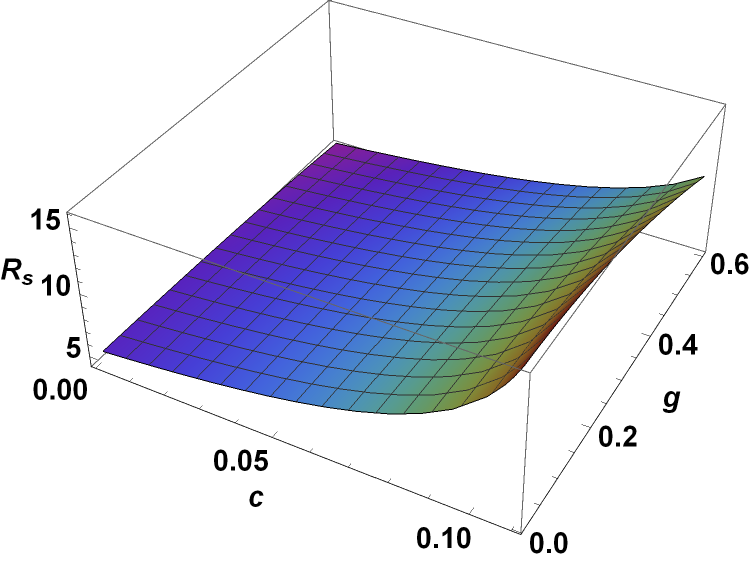}
    \caption{The profile of the shadow radius for various values of BH parameters $c$  and $g$ showing that the $R_{s}$ decreases with $g$ but increases with $c$. Here, $M=1$,  $\ell_p=100$.}
    \label{figa2a}
\end{figure}

Considering  the metric function (\ref{m1}), then the equation for $r_{ph}$ is 
\begin{equation}
\left(cr-2\right)\left(r^2+g^2\right)^{3}-4g^2r^4+6Mr^4\left(r^2+g^2\right)^{1/2}=0. \label{eps1}
\end{equation} 
The solution to this equation (\ref{eps1}) yields photon sphere. However, since this equation cannot be solved analytically, it can be solved numerically. The numerical photon sphere radius values for various MM charges  and QF parameter values are appended to table \ref{taba2}.
\begin{center}
\begin{tabular}{|c|c|c|c|c|c|}
 \hline 
\rowcolor{lightgray} \multicolumn{5}{|c|}{ $r_{ph}$ }
\\ \hline \rowcolor{lightgray}
$g$ & $c=0.01$ & $0.05$ & $0.1$ & $0.15$  \\ \hline
$0.2$ & $2.98468$ & $3.20269$ & $3.60476$ & $4.45692$ \\ 
$0.4$ & $2.7807$ & $2.99257$ & $3.37636$ & $4.1443$ \\ 
$0.6$ & $2.32539$ & $2.5393$ & $2.90672$ & $3.56538$ \\ 
 
 \hline
\end{tabular}
\captionof{table}{Numerical results for the photon sphere with various BH parameters,  the positive constant $c$ and the  MM charge $g$. Here $M=1$ and $\ell_p=100$.} \label{taba2}
\end{center}
Table shows that the effects of the QF parameter, $c$, and the MM charge, $g$, on photon radii differ. Increasing the QF parameter $c$ leads to an increase in photon radius, while increasing the MM charge $g$ decreases it. The three-dimensional visualizations in Figure  \ref{figa1a} provide a comprehensive illustration of these contrasting behaviors.\\
  Using the photon sphere radius we can found the shadow radius $R_s$ observed at infinity by an observer as \begin{equation}
    R_s=\frac{r_{ph}}{\sqrt{f(r_{ph})}} .\label{shadeq1}
\end{equation}
Again numerical solution is used in order to found the shadow radius. The following table \ref{taba3} lists the numerical shadow radius values for different MM charges and QF values. 
\begin{center}
\begin{tabular}{|c|c|c|c|c|c|}
 \hline 
\rowcolor{lightgray} \multicolumn{5}{|c|}{ $R_{s}$ }
\\ \hline \rowcolor{lightgray}
$g$ & $c=0$ & $0.01$ & $0.05$ & $0.1$ \\ \hline
$0.2$ & $5.11887$ & $5.36156$ & $6.76811$ & $11.9096$ \\ 
$0.4$ & $4.88986$ & $5.11669$ & $6.41367$ & $10.8378$ \\ 
$0.6$ & $4.40648$ & $4.60499$ & $5.70979$ & $9.05463$ \\ 
 
 \hline
\end{tabular}
\captionof{table}{Numerical results for the shadow radius with various BH parameters, the positive constant $c$ and the  MM charge $g$. Here $M=1$ and $\ell_p=100$.} \label{taba3}
\end{center}
Clearly, the shadow radius increases as QF parameter $c$ increases, but decreases as MM charge $g$ increases. Figure \ref{figa2a}  illustrates how parameters ($g,c$) affect the shadow radius. \\ To represent the actual shadow of the BH as seen from an observer's perspective, we introduce celestial coordinates, $X$ and $Y$
\begin{equation}
X=\lim_{r_{\mathrm{o}}\rightarrow \infty }\left( -r_{\mathrm{o}}^{2}\sin
\theta _{\mathrm{o}}\frac{d\varphi }{dr}\right) ,
\end{equation}%
\begin{equation}
Y=\lim_{r_{\mathrm{o}}\rightarrow \infty }\left( r_{\mathrm{o}}^{2}\frac{%
d\theta }{dr}\right) .
\end{equation}

\begin{figure}
    \centering
    \includegraphics[scale=0.6]{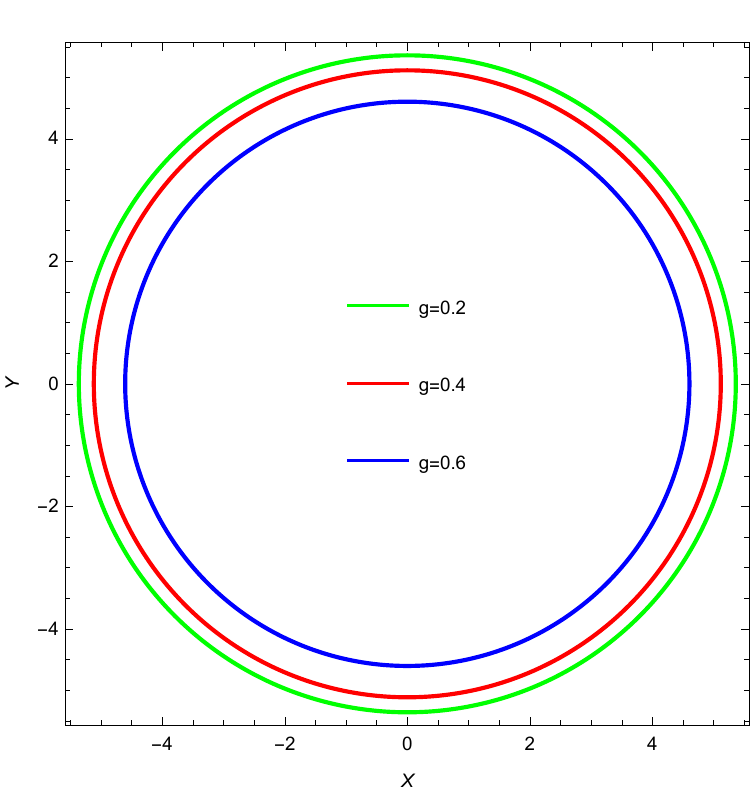} \includegraphics[scale=0.6]{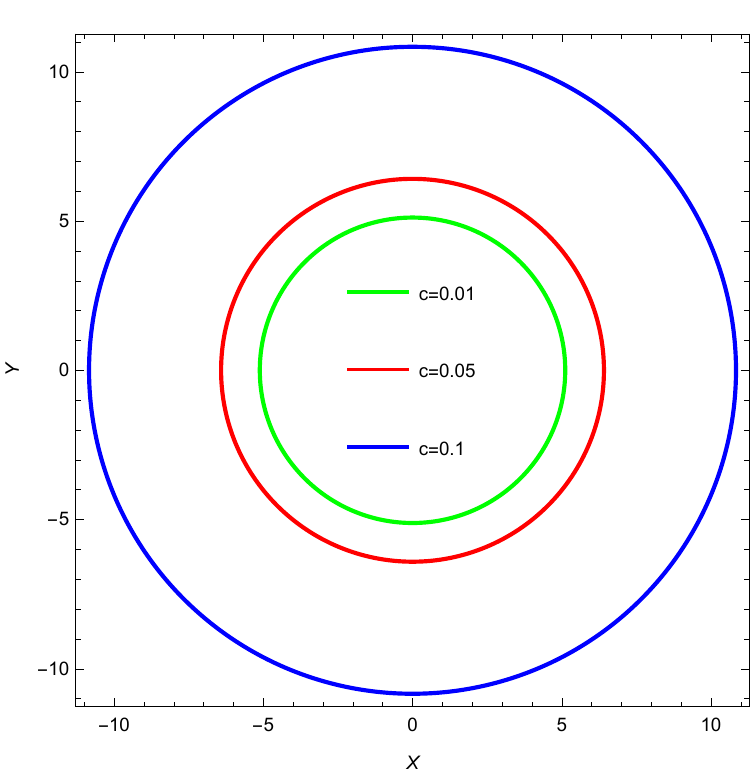}
    \caption{BH shadows  for different values of $g$  keeping  $c= 0.05$ (left) and for different values of $c$ keeping $g=0.4$ (right).}
    \label{ps25}
\end{figure}

For a static observer at large distance, i.e. at  $r_{\mathrm{o}}\rightarrow
\infty $ in the equatorial plane $\theta _{\mathrm{o}}=\pi /2$, the
celestial coordinates simplify to 
\begin{equation}
X^{2}+Y^{2}=R_s^{2}.
\end{equation}
Figure \ref{ps25} illustrates how the shadow changes with BH parameters. It demonstrates that the presence of quintessence causes the size of the shadow to expand while decreasing with the MM charge.

\section{Thermal properties of BH}\label{sec:6}

\begin{figure}[ht!]
\begin{center}
\includegraphics[scale=0.9]{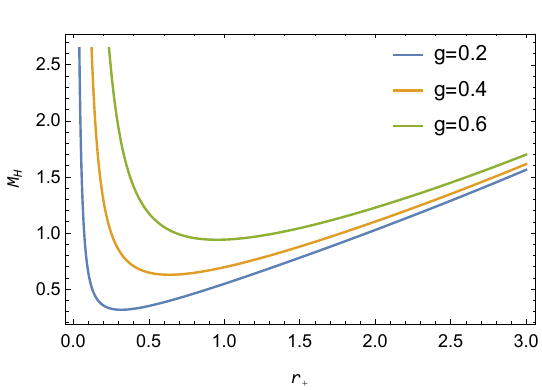}\quad
\includegraphics[scale=0.9]{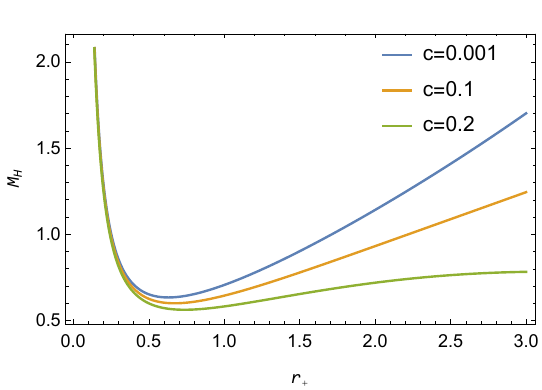}
\end{center}
\caption{The mass of ABG BH with QF showing the influence of the MM charge (left) and the normalization constant (right).  Here $\ell_p=100.$.}\label{figa1}
\end{figure}

\begin{figure}[ht!]
\begin{center}
\includegraphics[scale=0.9]{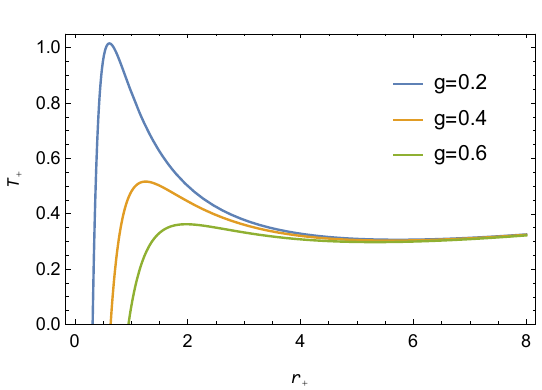}\quad
\includegraphics[scale=0.9]{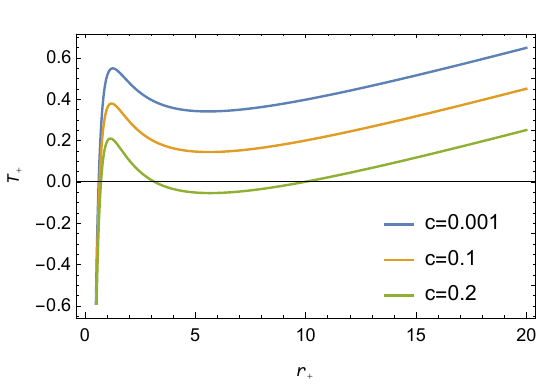}
\end{center}
\caption{The temperature of ABG BH with QF showing the influence of the MM charge (left) and the normalization constant (right).  Here $\ell_p=100$.}\label{figa2}
\end{figure}

In this section we will look at the thermodynamic quantities associated with the BH solution (\ref{m1}), which is defined by the parameters $M,\ell_p ,g$ and $c$. To serve the
 purpose, we discuss the thermodynamic parameters of mass, temperature, entropy, specific heat capacity and Gibbs free energy. \\ We begin by considering the mass of a BH, applying the horizon condition $f(r)=0$ at the horizon radius $r_{+}$.  Thus,
\begin{equation}
M=\frac{\left(
r_{+}^{2}+g^{2}\right)^{3/2}}{2r_+^2}\left(  1+\frac{r^2}{\ell^2_{p}}-cr+\frac{
r_{+}^{2}g^{2}}{
r_{+}^{2}+g^{2}}  
       \right).  \label{mass1}
\end{equation}%
In the limit, without QF ($c=0$) Eq. (\ref{mass1}) reduces to the mass of AdS ABG BH \cite{abg21} and when  $\ell_p \rightarrow{\infty}$ it  reduces to the mass of the ABG BH as,\begin{equation}
M=\frac{g^{4}+3\,r_{+}^{2}g^{2}+r_{+}^{4}}{2\,r_{+}^{2}\sqrt{r_{+}^{2}+g^{2}}}.  \label{mass11}
\end{equation}%
Let us examine the mass function by generating figure \ref{figa1}. The figure  illustrates that the mass function is affected by the MM charge and QF, but its characteristic features remain unchanged. Its mass reaches its lowest point after falling exponentially and nearing zero at a certain critical horizon radius. In contrast to the Schwarzschild BH, the BH's mass increases linearly with its radius. Furthermore, we can see that as $g$ increases, so does the mass of the BH. However, the mass decreases as the quintessence parameter $c$ grows.\\
The Ads ABG BH temperature, or Hawking temperature, can be calculated using $T_+=f'(r_+)/4 \pi$, thus   
\begin{equation}
T_{+}=\frac{1}{4\pi}\left[\frac{-2}{r_{+}}+ \left(  \frac{3g^{2}}{r_{+}^{2}+g^{2}}-2          \right)c+\frac{3r_{+}g^{4}(r_{+}^{2}+g^{2})+3r_+^5(r_{+}^{2}+\ell^{2}_p)+g^2\,r_+^3(5\,\ell^{2}_p+6\,r_+^2)}{\ell^2_{p}(r_{+}^{2}+g^{2})^3}\right].  \label{temp1}
\end{equation}
In the limit, ($c=0$) Eq. (\ref{temp1}) reduces to the Hawking temperature  of AdS ABG BH \cite{abg21} and  for $\ell_p \rightarrow{\infty}$, it reduces to the temperature of the ABG BH as, 
\begin{equation}
T_+=\frac{-2\,g^{6}-3\,r_{+}^{2}g^{4}-r_{+}^{4}g^{2}+r_{+}^{6}}{4\pi \,r_{+}\left(r_{+}^{2}+g^{2}\right)^3}.  \label{temp11}
\end{equation}
The temperature (\ref{temp1}) of AdS ABG BH with QF  is characterized by the parameters charge $g$, normalization constant $c$ and cosmological constant. In  Figure \ref{figa2}, we provide a plot of Hawking temperature vs horizon and analyse how the parameters ($g,c$) affect it. In all cases, the temperature of the BH rises abruptly to a maximum value at a certain horizon radius, then decreases exponentially with increasing horizon radius, and eventually rises as horizon radius approaches infinity. The maximum temperature of a BH decreases and shift right with increasing parameters $g$. 

Let us now apply the first law of thermodynamics to the other most significant thermodynamic number, the BH's entropy. The first law of thermodynamics ($dM=T_{+}dS_{+}+\Phi\,dg$) can be used to calculate the explicit form of entropy ($S_+$).    
This leads to 
\begin{equation}
S_{+}=\int \frac{1}{T_{+}}\frac{\partial M}{\partial r_{+}}dr_{+}=\pi\left(r_{+}-\frac{2g^2}{r_{+}}    \right) \sqrt{r_{+}^{2}+g^{2}}+3\,\pi\,g^{2}\log[r_{+}+\sqrt{r_{+}^{2}+g^{2}}].
\label{entr2}
\end{equation}%
In this case, it is evident that the entropy does not follow the area law. Using the first law of thermodynamics, we can also calculate the temperature based on entropy 

\begin{equation}
T_+=\frac{\partial M}{\partial S_+}=\frac{1}{4\pi}\left[\frac{-2}{r_{+}}+ \left(  \frac{3g^{2}}{r_{+}^{2}+g^{2}}-2          \right)c+\frac{3r_{+}g^{4}(r_{+}^{2}+g^{2})+3r_+^5(r_{+}^{2}+\ell^{2}_p)+g^2\,r_+^3(5\,\ell^{2}_p+6\,r_+^2)}{\ell^2_{p}(r_{+}^{2}+g^{2})^3}\right]. 
\end{equation}

The heat capacity is used to assess a BH's thermal stability as a thermodynamic system. Investigating the heat capacity allows us to evaluate if the thermodynamic system is locally stable or unstable. In fact, this can be performed by checking the sign of the heat capacity, which corresponds to a stable thermodynamic system with a positive sign, whereas an unstable state has a negative value.  In classical thermodynamics, the specific heat ($C_+$) is determined using the following standard equation: $C_{+}=\frac{dM}{dT_+}$ 

{\small 
\begin{equation}
    C_+=\frac{2\pi\,\ell^2_p\,r_{+}^{2}(r_{+}^{2}+g^{2})^4}{2g^8\ell^2_p-r_{+}^{8}(\ell^2_p-3r_{+}^{2})+3g^4r_{+}^{4}(7r_{+}^{2}+4\ell^2_p(1-cr_{+}))-g^6r_{+}^{2}(\ell^2_p(6cr_{+}-11)-9r_{+}^{2})-g^2r_{+}^{6}(\ell^2_p(6cr_{+}-8)-15r_{+}^{2})}. \label{heatc1}
\end{equation}
}

Specific heat capacity (\ref{heatc1}) depends on the MM charge $g$, the cosmological constant $\Lambda=-3/\ell^2_p$, and the constant $c$ for a particular parameter of the state $w=-2/3$. Without QF ($c=0$) Eq. (\ref{heatc1}) reduces to the heat capacity of AdS ABG BH \cite{abg21} and for $\ell_p \rightarrow{\infty}$ it  reduces to the specific heat capacity of the ABG BH as, 

{\begin{equation}
    C_+=\frac{2\pi\,r_{+}^{2}(r_{+}^{2}+g^{2})^4}{2g^8+11g^6r_{+}^{2}+12r_{+}^{4}+8g^2r_{+}^{6}-r_{+}^{8}}. \label{heatc2}
\end{equation}

Because the heat capacity equation  (\ref{heatc1}) is complex and difficult to interpret, it is displayed in Fig. \ref{fig11a} for various values of charge $g$ and a fixed values of $c=0.02$ and $\ell_p=100$. The plot reveals that the heat capacity diverges at the critical  points $r_2$ and $r_3$. According to figure, AdS ABG with QF is stable in the regions $r_+<r_1$ and $r_+>r_3$. Whereas the unstable region is $r_1<r_+<r_2$. This plot clearly shows that the AdS ABG BH with QF passes through two phase transitions: $r_2$  from the smaller stable BH to the larger unstable BH, and $r_3$  from the smaller unstable BH to the larger stable BH ($r_+>r_3$).  

\begin{figure}
    \centering
    \includegraphics[width=0.5\linewidth]{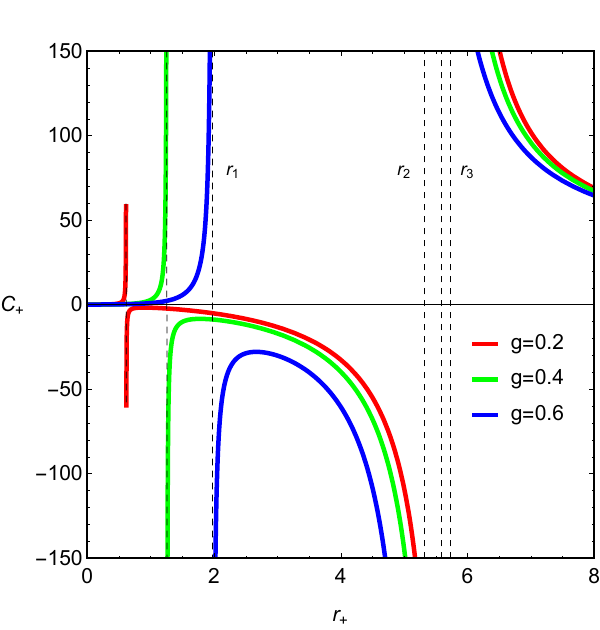}
    \caption{Variation of the specific heat capacity $ C_+$ as a function of horizon radius $r_+$ for various values of the charge $g$.
 Here $c=0.02, \ell_{p}=100$.}
    \label{fig11a}
\end{figure}

To investigate global thermodynamic stability, we now study the Gibbs free energy. In thermodynamics, Gibbs free energy is the maximum quantity of mechanical work that may be extracted from a system. It is mathematically defined as the following relation: $G_+=M_+-T_+S_+$. The Gibbs free energy can be expressed by using Eqs. (\ref{mass1}), (\ref{temp1}), and (\ref{entr2}): 
\begin{equation*}
    G_+=\frac{\left(
r_{+}^{2}+g^{2}\right)^{3/2}}{2\,r_+^2}\left(1+\frac{r^2_{+}}{\ell^2_{p}}-c\,r_{+}+\frac{
r_{+}^{2}\,g^{2}}{
r_{+}^{2}+g^{2}}  
       \right)-
\end{equation*} 

\begin{eqnarray}
      & &\left[\frac{-2}{r_{+}}+ \left(\frac{3\,g^{2}}{r_{+}^{2}+g^{2}}-2          \right)\,c+\frac{3\,r_{+}\,g^{4}\,(r_{+}^{2}+g^{2})+3\,r_+^5\,(r_{+}^{2}+\ell^{2}_p)+g^2\,r_+^3(5\,\ell^{2}_p+6\,r_+^2)}{\ell^2_{p}(r_{+}^{2}+g^{2})^3}\right] \times \nonumber\\ &&\left\{\left(r_{+}-\frac{2\,g^2}{r_{+}}    \right) \sqrt{r_{+}^{2}+g^{2}}+3\,\pi\,g^{2}\log\left[r_{+}+\sqrt{r_{+}^{2}+g^{2}}\right]   \right\}. \label{gib1}
\end{eqnarray}

If the QF parameter $c$ is set to zero, the Gibbs free energy expression simplifies to the AdS ABG BH free energy. The expression (\ref{gib1}) clearly demonstrates that the QF affects the Gibbs free energy. Figure 10 shows  Gibbs free energy plots to visualize the impact of QF $c$ and the MM cgarge $g$ parameters. Figure \ref{figa5} shows that the QF parameter $c$ and MM cgarge $Q$ have opposite effects on the Gibbs free energy ($G_+$). Increasing the parameter $g$ leads to a higher Gibbs free energy, which has a greater impact on smaller BHs. Furthermore, we discover a global minimum ($r_{min}$) and a global maximum ($r_{max}$), which correspond to the Hawking temperature's extreme points. At these important points, the state of free energy changes. Specifically, beyond the minimum radius $r_{min}$, the free energy increases with increasing horizon radius $r_+$, peaking at $r_{max}$. After this optimum, the free energy continues to decrease as the horizon radius rises.
\begin{figure}[ht!]
\begin{center}
\includegraphics[scale=0.9]{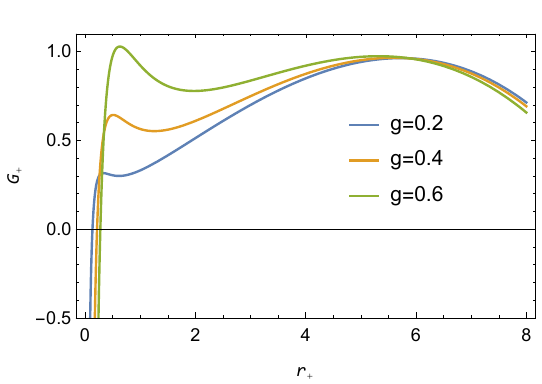}\quad
\includegraphics[scale=0.9]{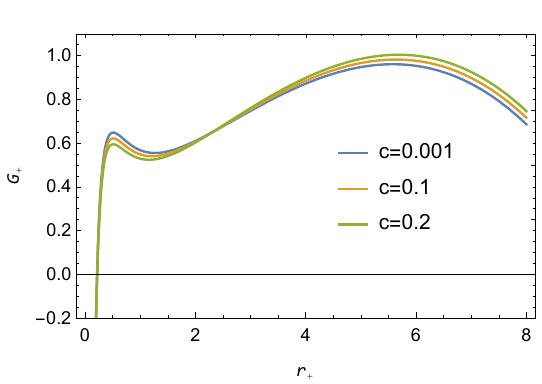}
\end{center}
\caption{The Helmholtz energy of the  ABG BH with QF showing the influence of the MM charge (left) and the normalization constant (right).  Here $\ell_p=100$.}\label{figa5}
\end{figure}

\section{Summary and Conclusions}\label{sec:7}

In this study, we comprehensively investigated the ABG BH solution in AdS space-time surrounded by a QF with state parameter $w = -2/3$. We derived an exact analytical solution to the gravitational field equations for this unique configuration, characterized by parameters including the BH mass $M$, the magnetic monopole charge $g$, the cosmological constant related to AdS length $\ell_p = \sqrt{-3/\Lambda}$, and the QF parameter $c$. The metric function in Eq.~(\ref{m1}) was shown to reduce to some known BH solutions in appropriate limits, demonstrating the robustness of our derived solution.

Our analysis of geodesic motion showed significant effects both the ABG parameter $g$ and the quintessence parameter $c$ on the trajectories of test particles. For null geodesics, we observed that the effective potential, presented in Eq.~(\ref{pp6}), exhibited a distinct behavior compared to the standard BH solution. As illustrated in Fig.~\ref{fig:null-potential}, increasing the ABG parameter $g$ led to a decrease in the potential barrier height, while increasing the parameter $c$ modified the asymptotic behavior of the potential. These modifications directly influenced the stability and characteristics of circular photon orbits, which was evident from our calculations of the photon sphere radius as shown in Eq.~(\ref{pp9}). The force acting on photon particle, expressed in Eq.~(\ref{pp10}), demonstrated an intricate dependence on both $g$ and $c$. Fig.~\ref{fig:force} showed that the ABG parameter $g$ significantly altered the repulsive-attractive transition points of the force, while the parameter $c$ primarily affected the long-range behavior. This finding has important implications for understanding the gravitational lensing effects of such BH. For massive particle, our analysis of time-like geodesics showed equally interesting results. The effective potential given by Eq.~(\ref{qq1}) showed that stable circular orbits were possible under specific parameter conditions. The calculated energy and angular momentum of time-like particles in circular orbits, as presented in Eqs.~(\ref{qq3})-(\ref{qq4}), exhibited a complex dependence on both $g$ and $c$. Fig.~\ref{fig:energy} demonstrated that the energy profiles were significantly altered with changing these parameters, affecting the binding energy of orbiting particles. We then extended our analysis to the geodesic precession frequency, a phenomenon that has significant observational implications. Eq.~(\ref{qq8}) provided the mathematical expression for this frequency, and Fig.~\ref{fig:frequency} illustrated how both $g$ and $c$ modified the precession rates compared to standard BH solutions. This finding could potentially be tested against observed precession rates in strong gravitational fields. Moreover, the circular speed of time-like particles, calculated in Eq.~(\ref{qq13}), showed distinct variations with both $g$ and $c$, as visualized in Fig.~\ref{fig:speed}. This result has implications for understanding galactic rotation curves and could potentially connect to dark matter phenomenology. Our investigation of scalar perturbations produced the effective potential given by Eq.~(\ref{kk6}), which governs the propagation of massless scalar fields in this space-time. The structure of this potential indicated that the ABG parameter $g$ and the QF parameter $c$ significantly influenced the scattering properties of scalar waves. The BH shadow, a directly observable feature with current astronomical technology, was thoroughly analyzed. We determined the photon sphere radius through Eq.~(\ref{eps1}) and calculated the shadow radius as presented in Eq.~(\ref{shadeq1}). Our numerical results, summarized in Tables \ref{taba2} and \ref{taba3}, showed that the shadow radius increased with increasing $c$ but decreased with increasing $g$. Fig.~\ref{ps25} provided a visual representation of these shadows, demonstrating how the observable silhouette of the BH would appear to distant observers. This result is particularly significant as it provides a potential observational test for distinguishing these BH solutions from standard ones.

In addition, the thermodynamic properties of the ABG BH in AdS space-time with QF showed perhaps the most intricate behavior. We derived expressions for the BH mass (\ref{mass1}), the Hawking temperature (\ref{temp1}), and entropy (\ref{entr2}) of the BH. Fig.~\ref{figa2} illustrated the temperature profile as a function of horizon radius, showing characteristic peaks that were shifted by both $g$ and $c$. The entropy deviated from the standard area law, reflecting the modified gravitational theory underlying this solution. The specific heat capacity, given by Eq.~(\ref{heatc1}), exhibited divergences at critical points as shown in Fig.~\ref{fig11a}, indicating phase transitions in the thermodynamic behavior of the BH. We identified regions of thermodynamic stability ($r_+ < r_1$ and $r_+ > r_3$) and instability ($r_1 < r_+ < r_2$), demonstrating that the ABG BH with QF undergoes two distinct phase transitions. This behavior is significantly different from standard AdS BH and reflects the complex interplay between the ABG parameter, the QF, and the AdS curvature. Finally, our analysis of the Gibbs free energy, expressed in Eq.~(\ref{gib1}) and visualized in Fig.~\ref{figa5}, completed the thermodynamic picture. The free energy profile showed characteristic global minimum and maximum points corresponding to the temperature extrema, with significant modifications due to both $g$ and $c$. The global thermodynamic stability was found to be parameter-dependent, with transitions between different phases occurring at critical points.

Throughout our analysis, we consistently observed that the ABG parameter $g$ and the QF parameter $c$ had opposing effects on most physical quantities. Increasing $g$ typically led to behaviors closer to standard BHs, while increasing $c$ introduced more exotic features. The AdS curvature parameter $\ell_p$ primarily affected the asymptotic behavior and large-scale properties of the solution. For future research, several promising directions emerge from this study. First, the stability analysis of the ABG BH in AdS spacetime with QF could be extended to include electromagnetic and gravitational perturbations, providing a more complete picture of the solution's stability. Second, the calculation of quasinormal modes would offer insights into the gravitational wave signatures of such BH, potentially providing observational tests. Third, the holographic interpretation of this solution in the context of the AdS/CFT correspondence would be valuable for understanding the dual field theory and its phase structure. These future investigations will be in our future-works agenda.

{\small

\section*{Data Availability Statement}

No new data were generated or analyzed in this study.

\section*{Conflicts of Interest}

There is no conflict of interests.

\section*{Funding Statement}

No fund has received for this paper.

\section*{Code/Software}

No software/Coder were used in this study.

\section*{Acknowledgments}

F.A. acknowledges the Inter University Centre for Astronomy and Astrophysics (IUCAA), Pune, India for granting visiting associateship. \.{I}.~S. thanks EMU, T\"{U}B\.{I}TAK, ANKOS, and SCOAP3 for funding and acknowledges the networking support from COST Actions CA22113, CA21106, and CA23130.
}

\end{document}